\documentclass[twocolumn]{pasj00}
\SetRunningHead{TSUMURA et al.}{Zodiacal Light Spectrum with AKARI IRC} 
\Received{2013 January 30}
\Accepted{2013 June 5}
\Published{}         

\newcommand{\nw}{nWm$^{-2}$sr$^{-1}$}

\begin{document}
\title{Low-Resolution Spectrum of the Zodiacal Light with AKARI InfraRed Camera} 
\author{Kohji \textsc{Tsumura}\altaffilmark{1}, Toshio \textsc{Matsumoto}\altaffilmark{1,2}, Shuji \textsc{Matsuura}\altaffilmark{1}, Jeonghyun \textsc{Pyo}\altaffilmark{3}, Itsuki \textsc{Sakon}\altaffilmark{4}, and Takehiko \textsc{Wada}\altaffilmark{1}}
\altaffiltext{1}{Department of Space Astronomy and Astrophysics, Institute of Space and Astronautical Science, Japan Aerospace Exploration Agency, 3-1-1 Yoshinodai, Chuo-ku, Sagamihara, Kanagawa 252-5210}
\altaffiltext{2}{Institute of Astronomy and Astrophysics, Academia Sinica, No.1, Roosevelt Rd, Sec. 4, Taipei 10617, Taiwan, R.O.C.}
\altaffiltext{3}{Korea Astronomy and Space Science Institute, Daejeong 305-348, Republic of Korea}
\altaffiltext{4}{Department of Astronomy, Graduate School of Science, The University of Tokyo, Hongo 7-3-1, Bunkyo-ku, Tokyo 113-0033}
\email{tsumura@ir.isas.jaxa.jp}
\maketitle

\begin{abstract}
We present the near- and mid-infrared zodiacal light spectrum obtained with the AKARI Infra-Red Camera (IRC).
A catalog of 278 spectra of the diffuse sky covering a wide range of Galactic and ecliptic latitudes was constructed.
The wavelength range of this catalog is 1.8-5.3 $\mu$m with wavelength resolution of $\lambda /\Delta \lambda \sim 20$.
Advanced reduction methods specialized for the slit spectroscopy of diffuse sky spectra are developed for constructing the spectral catalog.
Based on the comparison analysis of the spectra collected in different seasons and ecliptic latitudes,
we confirmed that the spectral shape of the scattered component and the thermal emission component of the zodiacal light in our wavelength range does not show any dependence on location and time, 
but relative brightness between them varies with location.
We also confirmed that the color temperature of the zodiacal emission at 3-5 $\mu$m is 300$\pm $10 K at any ecliptic latitude.
This emission is expected to be originated from sub-micron dust particles in the interplanetary space.
\end{abstract}
 
\section{Introduction}
The astrophysical sky brightness is dominated by the zodiacal light (ZL) which comprises scattered sunlight by interplanetary dust (IPD) in the optical and near-infrared (NIR), 
and thermal zodiacal emission (ZE) from the same IPD in the mid-infrared (MIR) or longer\footnote{Sometimes the term ZL indicates only the scattered component to distinguish it from ZE. However, the term ZL indicates both scattered and thermal components in this paper, while the term ZE indicates only the thermal emission component.}.
Historically, extensive ZL observations in the scattered sunlight regime were conducted from ground-based observations at high altitude sites in the 1960's and 1970's \citep{Levasseur80}.
However, these were restricted to optical wavelengths because atmospheric thermal emission and OH airglow are much brighter than the ZL at NIR and MIR. 
Space based platforms get rid of contaminations and provide precise measurements of the diffuse sky brightness at NIR and MIR \citep{Murdock85, Berriman94, Matsuura95, Matsumoto96}.
Based on measurements from the COBE/DIRBE, 
physical models for the IPD distribution have been constructed \citep{Kelsall98, Wright98} to estimate the zodiacal foreground for measurements of the Extragalactic Background Light (EBL) \citep{Wright98, Hauser98, Matsumoto05, Matsumoto2013}.
Three major components are known in the DIRBE IPD distribution model \citep{Kelsall98}; (1) a smooth cloud which is dominant component of ZL ($>90\%$), 
(2) three asteroidal dust bands found by IRAS \citep{Low84} whose parent bodies were identified as being associated with certain asteroid families \citep{Dermott84, Nesvorny03},
and (3) a circumsolar ring which was formed by dusts resonantly trapped by the Earth \citep{Reach1995}.
Because materials likely to be present in the IPD have distinct spectral features at NIR and MIR, spectroscopic measurement of ZL helps to determine the composition of IPD particles.
For example, silicate features were found in the ZL spectrum at NIR by IRTS \citep{Matsumoto96} and CIBER \citep{Tsumura10}, and at MIR by IRTS \citep{Ootsubo98}, ISO \citep{Reach03} and AKARI \citep{Ootsubo09}.

\begin{figure*}
  \begin{center}
    \FigureFile(160mm,100mm){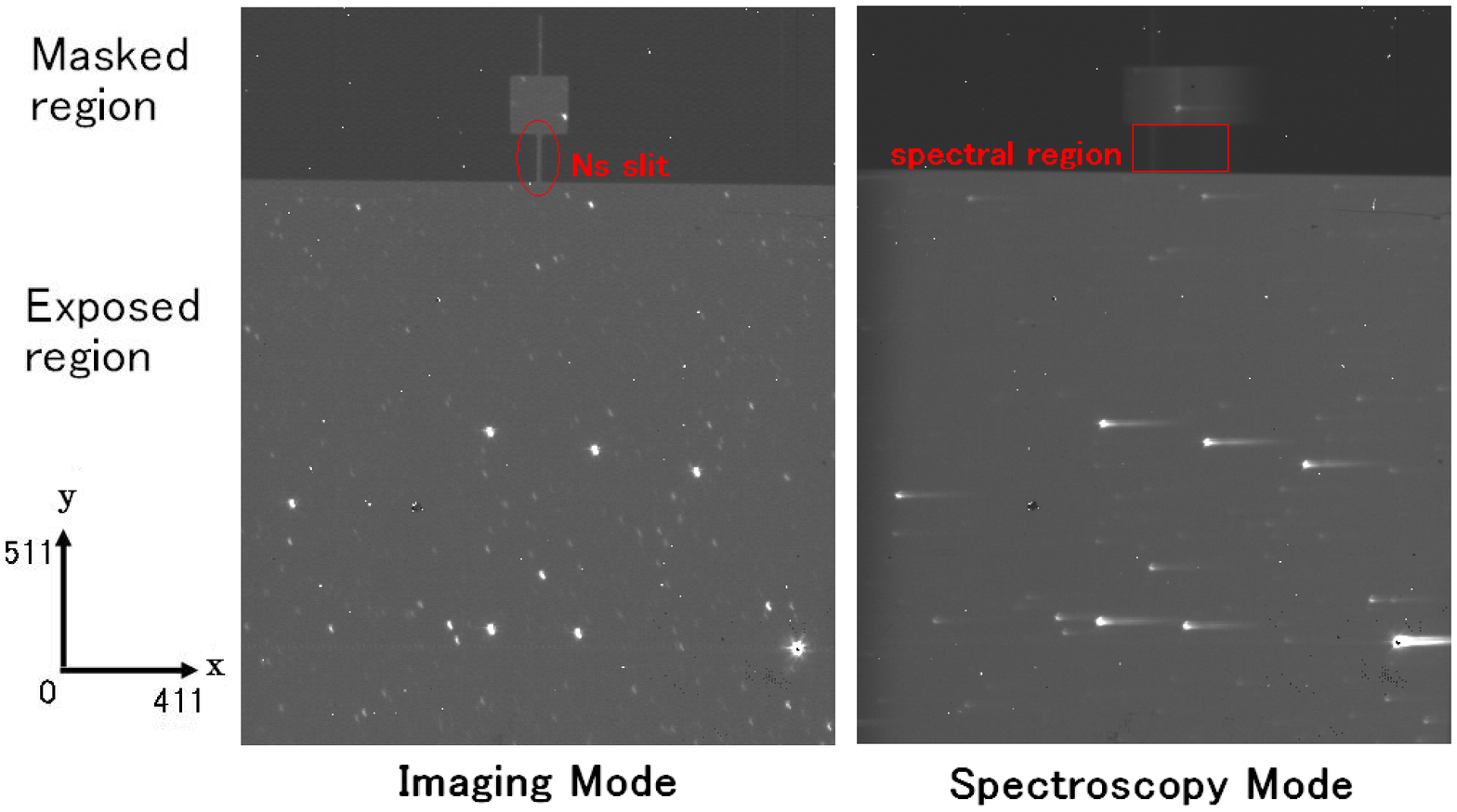}
  \end{center}
  \caption{An example image taken with low-resolution spectroscopic mode (NP) of AKARI IRC NIR channel (right) and the corresponding reference image taken with the imaging mode (left).
           A large part of the array is devoted as an exposed region for taking astronomical images.
           Three kinds of slits are also prepared at the top of the array, and both side of the slits are masked for monitoring dark currents during astronomical observations.
           In the spectroscopy mode, wavelength increases from left to right in this image.
            In this study, spectral data in the "spectral region" from the Ns slit were used.}
  \label{image}
\end{figure*}

In this paper, we describe ZL spectrum at 1.8-5.3 $\mu$m obtained with the low-resolution prism spectroscopy mode of the AKARI Infra-Red Camera (IRC) NIR channel.
The ZL spectrum has a local minimum in this wavelength region because the contributions from scattered component and thermal emission component from IPD are about equal at around 3.5 $\mu$m.
For this purpose, we constructed a low-resolution ($\lambda /\Delta \lambda \sim 20$) spectral catalog of diffuse sky spectrum in 1.8-5.3 $\mu$m wavelength region with IRC spanning wide range of ecliptic and Galactic coordinates.
This catalog of diffuse sky spectra is available at ISAS/JAXA\footnote{http://www.ir.isas.jaxa.jp/AKARI/Archive/Catalogues/ IRC\_\,diffuse\_\,spec/}.
There are two companion papers describing the spectrum of the infrared diffuse sky; 
the Diffuse Galactic Light (DGL) is described in \citet{Tsumura2013b} (hereafter Paper II)
and EBL is described in \citet{Tsumura2013c} (hereafter Paper III) 
in which the foregrounds described in Paper II and this paper (Paper I) are subtracted.

\section{Data Reduction}
\subsection{Dataset}
AKARI is the first Japanese infrared astronomical satellite launched in February 2006, equipped with a cryogenically cooled telescope of 68.5 cm aperture diameter \citep{Murakami07}.
IRC is one of two astronomical instruments of AKARI, and it covers 1.8-5.3 $\mu$m wavelength region with a 512$\times $412 InSb detector array in the NIR channel\footnote{IRC has two other channels covering 5.8-14.1 $\mu$m in the MIR-S channel and 12.4-26.5 $\mu$m in the MIR-L channel.} \citep{Onaka07}.
Replacing the imaging filters with a prism (called "NP") on the filter wheels, it provides low-resolution ($\lambda /\Delta \lambda \sim 20$) slit-less spectroscopy in its exposed region for point sources,
and slit spectroscopy for diffuse radiation\footnote{High-resolution spectroscopy ($\lambda /\Delta \lambda \sim 120$) with a grism (called "NG") is also available.} \citep{Ohyama2007}.
As shown in Figure \ref{image}, the slit for the NIR channel consists of three parts of different width, and a slit labeled 'Ns' (closest to the exposed region) 
with 5 arcsec width by 0.8 arcmin length, equivalent to 2.8$\times $27.2 pixels, are used in this study.
The rectangle region of 65$\times $27 pixels is used for diffuse sky spectrum, and $x$-direction corresponds to wavelength and $y$-direction corresponds to space.
In this paper, we call this rectangle region of 65$\times $27 pixels as "spectral region".
Performance of IRC was best during the cold mission\footnote{In this paper, the cold mission means the performance verification phase, phase-1, and phase-2 in official nomenclatures.}
before the exhaustion of liquid-helium for 550 days.
Temperature of IRC was stable and kept less than 7 K during the cold mission \citep{Nakagawa07}.
One of AKARI's advantages over Spitzer for observation of diffuse sky is a cold shutter.
AKARI equips the black position in its filter wheel so it can measure the dark current for the absolute photometry of the diffuse radiation, 
while the cold shutter of Spitzer has not been operated in the orbit \citep{Fazio2004}. 

\begin{figure*}
  \begin{center}
    \FigureFile(160mm,100mm){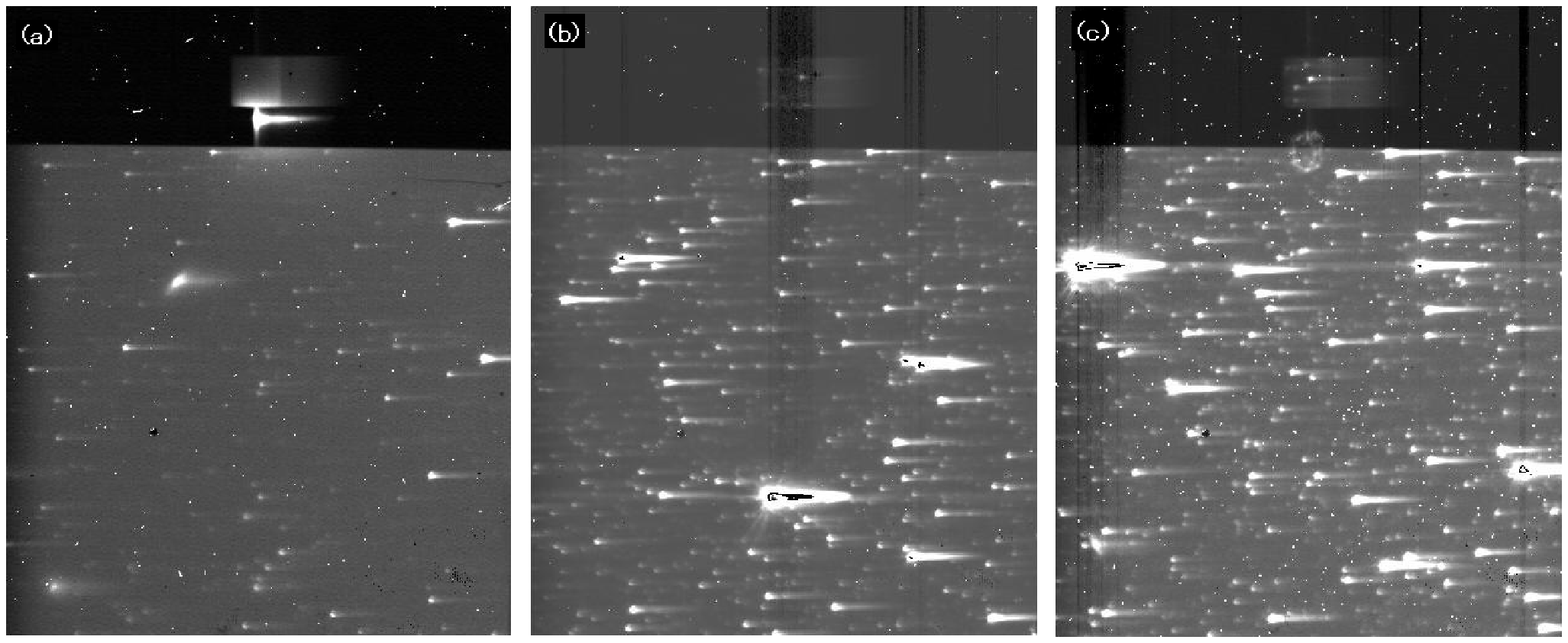}
  \end{center}
  \caption{Examples of excluded images in our data set. (a) A bright object is on the slit and no pixels can be used for diffuse sky spectrum.  (b) The spectral region is contaminated by column pull down from a bright source at the exposed region.
               (c) The spectral region is contaminated by a ghost image from a bright source at the exposed region.}
  \label{bads}
\end{figure*}

\begin{figure*}
  \begin{center}
    \FigureFile(160mm,100mm){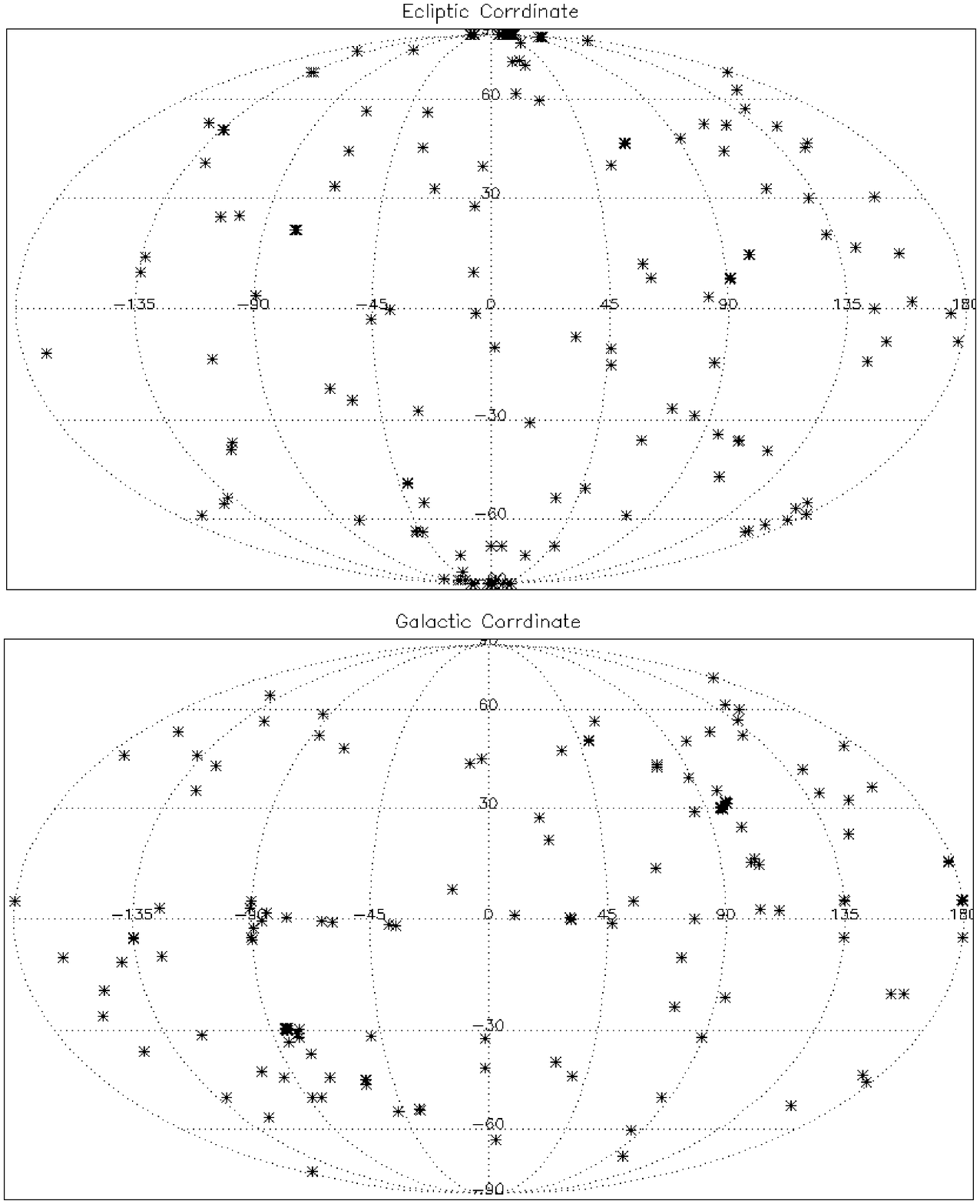}
  \end{center}
  \caption{Distribution of the pointed data in this study on the ecliptic (top) and Galactic (bottom) coordinates.}
  \label{Data_distribution}
\end{figure*}

\begin{figure}
  \begin{center}
    \FigureFile(80mm,50mm){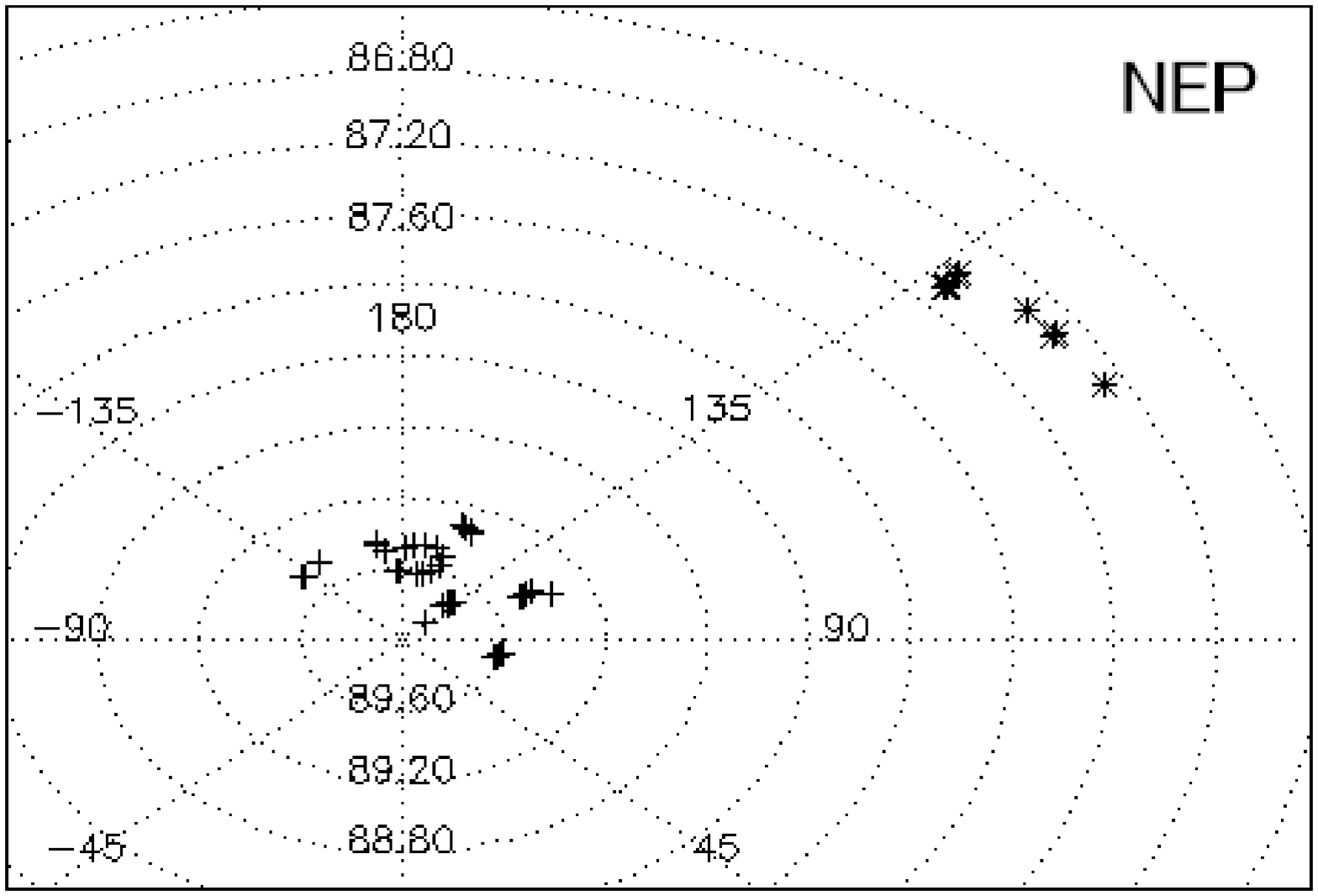}
  \end{center}
  \caption{Distribution of the pointed data at NEP region. Crosses indicate AKARI NEP field and asterisks indicate IRAC dark field.}
  \label{NEP_distribution}
\end{figure}

Dataset used in this study was obtained via AKARI Catalogue Archive Server (CAS) \citep{Yamauchi2011}.
First, 349 pointed data were selected on the condition that (1) it includes prism spectroscopy data, and 
(2) it was taken in the cold mission without any Earthshine contamination (from September 2006 to May 2007, see \citet{Pyo2010} for the Earthshine contamination).
Some data were excluded because of stray light from very bright sources in the exposed region as shown in Figure \ref{bads}.
The data in which some astronomical objects (eg. Orion nebula or Magellanic clouds) are on the slit were also excluded because our target is general interstellar/intergalactic region for the diffuse sky spectra.
In final, 278 pointed data were obtained as shown in Table \ref{DataTable}, Figure \ref{Data_distribution}, and Figure \ref{NEP_distribution}.
The data points are distributed over wide range of ecliptic and Galactic coordinates as shown in Figure \ref{Data_distribution} although they are not perfectly non-biased.
Because of the AKARI's Sun-synchronous polar orbit, North Ecliptic Pole (NEP) region, including  AKARI NEP field \citep{Matsuhara2006} and IRAC dark field   \citep{Krick2009, Krick2012}, was frequently observed in different seasons as shown in Figure \ref{NEP_distribution}.

\begin{table}
  \caption{Data number and position in this study}\label{DataTable}
  \begin{center}
    \begin{tabular}{lc}
      \hline
      Position & Number \\
      \hline
      AKARI NEP field & 80 \\
      IRAC dark field  & 38 \\
      $b > 5^{\circ }$ (w/o NEP nor IRAC fields) & 56 \\
      Galactic plane ($\mid b \mid < 5^{\circ }$) & 35 \\
      $b < -5^{\circ }$ & 69 \\
      \hline
      total & 278 \\
      \hline
    \end{tabular}
  \end{center}
\end{table}

\subsection{Dark current subtraction}
In this study, we adopted special data reduction besides the public tool kit \citep{Lorente2008}.
First, we developed a new dark current estimation method specialized for the diffuse sky analysis, 
in which the dark current of each pixel was estimated from the averaged dark current at the masked region.
This method provides a more reliable estimate of the dark current suffering from an after effect due to strong cosmic ray hits occurring in the South Atlantic Anomaly (SAA) region.
Typical value of dark current in the spectral region is 5-50 ADU, and error due to dark current subtraction by this method is reduced to about 1 ADU equivalent to 3 \nw\ at 2 $\mu$m.
See \citet{TsumuraWada2011} for more details.

\subsection{Linearity correction}
It can be assumed that the detector behaves linearly below 5000 ADU \citep{Lorente2008}.
Since the signals of diffuse sky even at the Galactic plane in this study is less than 2000 ADU, no linearity correction is required for our analysis.

\subsection{Flat field correction}
Since the flat fielding procedure for slit spectroscopy provided in the public tool kit \citep{Lorente2008} is not sufficient, we constructed a new flat field specialized for our purpose from our dataset.
Our dataset includes 80 pointed data at the AKARI NEP field (see Table \ref{DataTable}) where ZL dominates the sky brightness.
Although the absolute brightness of ZL varies with seasons, its spectral shape is almost the same as shown in the latter part in this paper.
Therefore, the new flat field was constructed by NEP data.

\begin{figure}
  \begin{center}
    \FigureFile(80mm,50mm){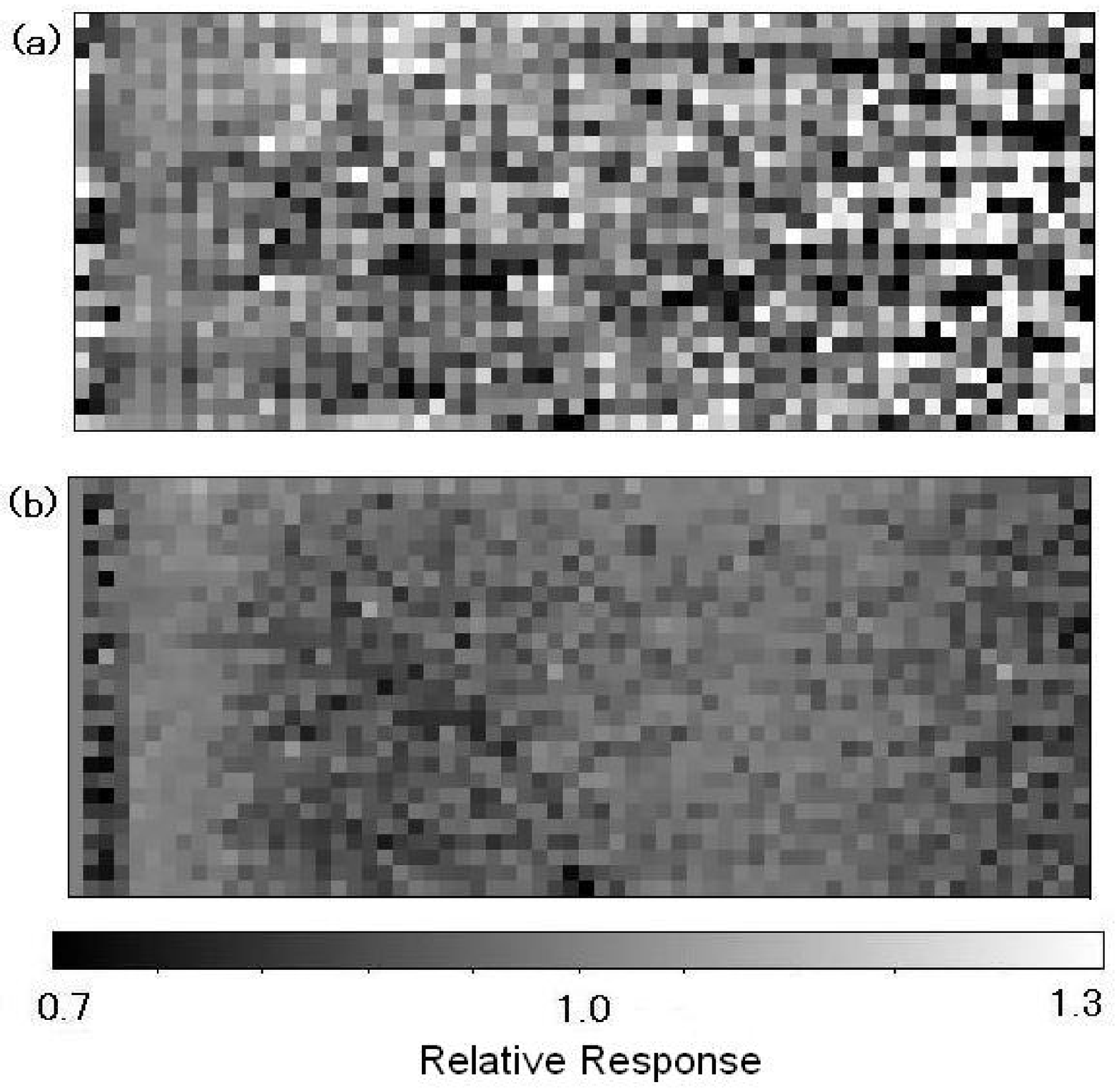}
  \end{center}
  \caption{Flat field image at the spectral region provided in the public tool kit (a) and that obtained by our method (b).
               A large scale distribution of sensitivity is reproduced in our new flat field with less deviation.} 
  \label{flat}
\end{figure}

By assuming that the diffuse sky brightness is constant along the slit length ($y$-direction, 0.8 arcmin), relative sensitivity of pixels along the slit for all given $x$ columns in the spectral region of a given image can be corrected by comparing to the averaged signal,
and a flat field was obtained by averaging the all 80 spectral images.
Figure \ref{flat} compares the resultant our flat field image with the existing one in the public tool kit.
A large scale distribution of sensitivity is reproduced in our new flat field with less deviation.
This flat field correction reduces the errors by 5\% comparing to the existing old flat field.  
Note that relative sensitivity along the x-direction cannot be corrected in our method since no light source with flat (white) spectrum is available. 
In addition, it cannot correct column pull downs owing to hot pixels fixed in the detector array. 

\subsection{Masking}
Field stars falling on the slit should be masked to derive the diffuse sky spectrum.
Data package of each field includes several raw frames depending on Astronomical Observation Template (AOT; see \citet{Onaka07} for details of AOT),
and position of stars are slightly different among these raw frames due to the attitude instability of the satellite during pointed observations.
Therefore, stars should be masked on each raw frames.
Spectra of point sources appear as horizontal lines on the spectral images, which were recognized and masked.
In this procedure, stars brighter than $m_K(\textrm{Vega}) = 19$ were detected on the slit and masked.
It was confirmed that the brightness due to unresolved Galactic stars under this detection limit is negligible by a Milky Way star counts model, TRILEGAL \citep{Girardi2005}.
Cumulative brightness contributed by unresolved galaxies can be estimated by the deep galaxy counts, being $<$4 \nw\ at K band in the case of limiting magnitude of $m_K = 19$ \citep{Keenan10}.

\begin{figure*}
  \begin{center}
    \FigureFile(120mm,80mm){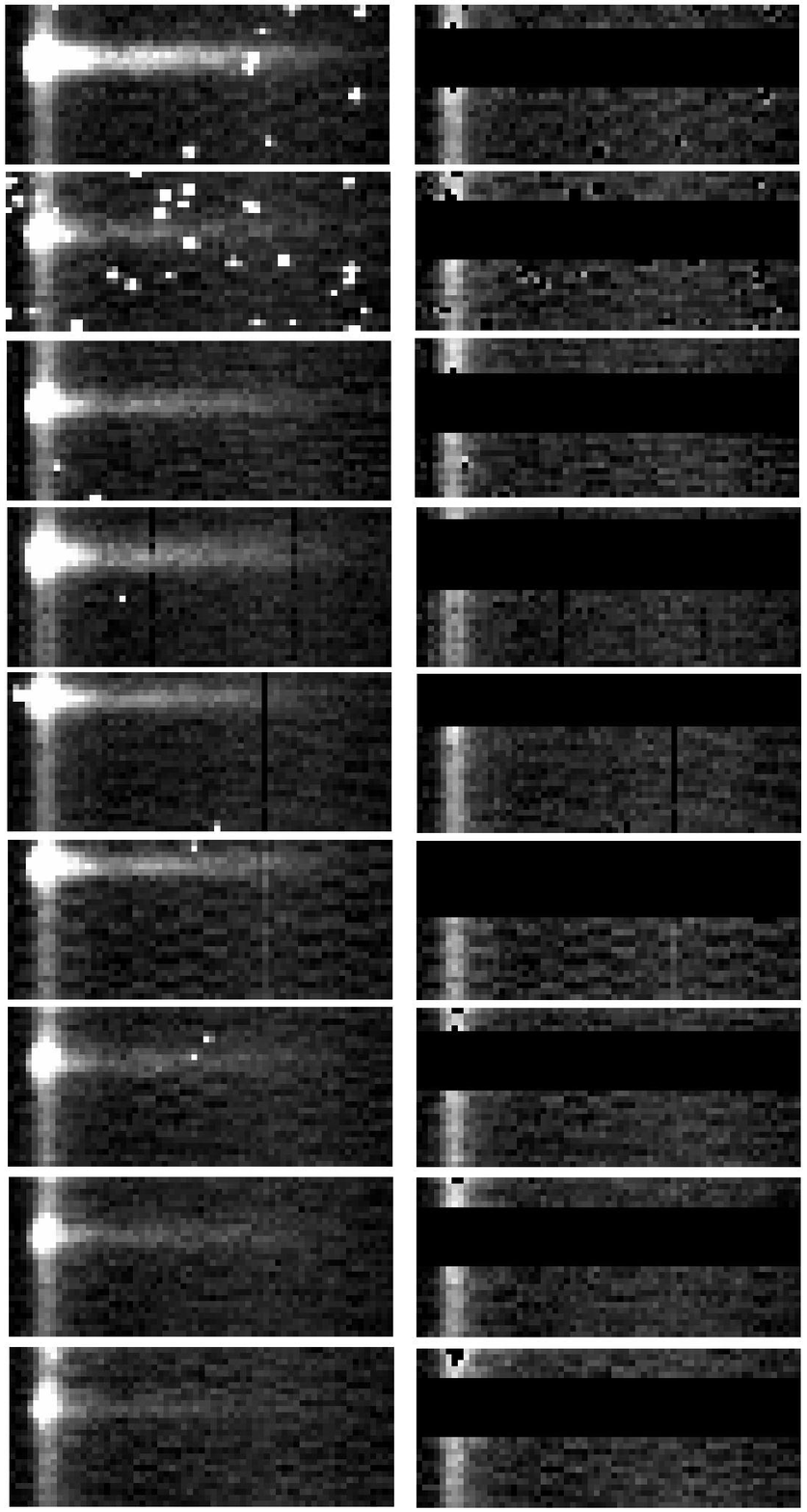}
  \end{center}
   \caption{Spectral images at the spectral region as an example of masking.  Left images are raw spectral images and right images are masked spectral images.
              This pointed data have nine raw frames and the detected star moves on the slit due to the attitude instability of the satellite. 
               Stars and hot pixels are masked as shown in right images.
               Although "week hot pixels" remain in the data, they are removed by 3$\sigma $ clipping procedure when these spectral images are averaged to derive a one-dimensional spectrum.}
  \label{mask}
\end{figure*}

In addition to the foreground field stars, hot pixels caused by cosmic ray hits on the detector array should be also masked on each raw frame.
Such hot pixels can be found as outliers and masked.  
Figure \ref{mask} shows an example of spectral images with masking of detected stars and hot pixels. 
Remaining "week hot pixels" in Figure \ref{mask} are removed with 3$\sigma $ clipping to derive a one-dimensional spectrum.

\subsection{Baseline subtraction}
After the reductions of dark current subtraction, flat field correction, and masking, 
signals on the spectral region were averaged along the $y$-direction at each $x$ bin with 3$\sigma $ clipping to removing the remaining week hot pixels.
In this procedure, the one-dimensional spectra of the diffuse sky in pixel-ADU units were obtained,
and the statistic error is evaluated by the dispersion along the $y$-direction.

\begin{figure}
  \begin{center}
    \FigureFile(80mm,50mm){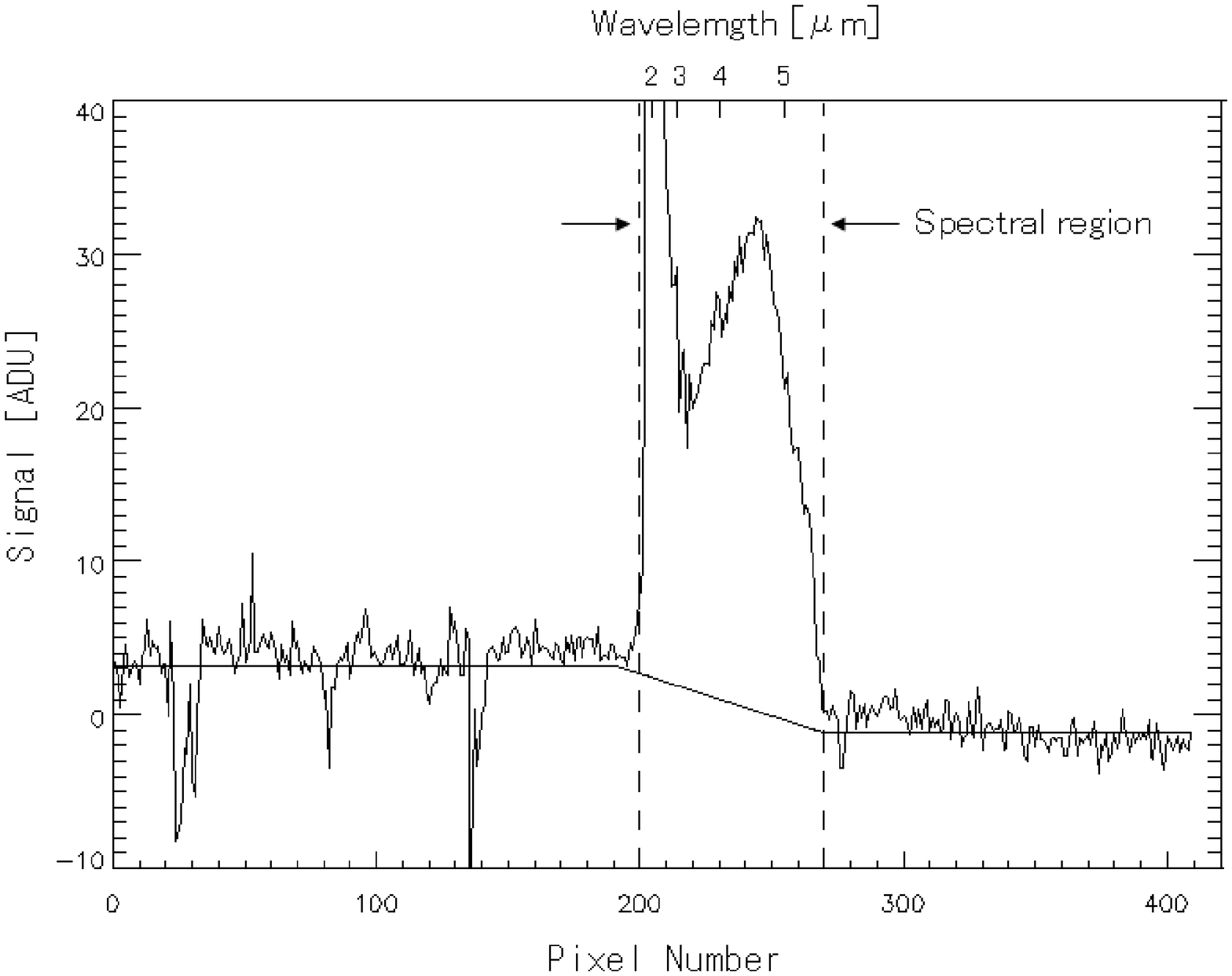}
  \end{center}
  \caption{An example of one-dimensional spectrum of diffuse sky.  Signals from the sky appears at pixels of $x=200-270$ shown by the broken lines, and others are masked region.
              Left-side masked region has non-zero counts because of scatter light from the exposed region, so it was modeled practically and subtracted.
              Small negative values at $x > 300$ is caused by the systematic error owing to the dark frame subtraction, 
              and a small excess emission at $x=280-300$ is caused by a contamination from the exposed region.} 
  \label{baseline}
\end{figure}

An example of obtained one-dimensional spectra is shown in Figure \ref{baseline}.
Signals from the sky appears at pixels of $x=200-270$, and others are masked region.
Although signals at the masked region should be zero because dark current was subtracted, left-side masked region has non-zero signals.
It is thought to be owing to scattered light from the exposed region because it is correlated weakly with the sky brightness, 
but an exact reason is still unknown.
In this study, we modeled this excess component practically by a simple interpolation and subtracted it from the spectra as shown in Figure \ref{baseline}.
Statistical dispersion from this model is $<2$ \% of the sky brightness, which is included in the statistical uncertainty.

\subsection{Pixel to Wavelength Relation}
Next, we need to convert pixel position ($x$-axis) to wavelength, and ADU to physical unit in surface brightness. 
As described in \citet{Lorente2008}, the pixel position $x$ for a given wavelength $\lambda $ [$\mu$m] is expressed in a 2nd-order polynomial equation as,
\begin{equation} x = \alpha \lambda ^2 +\beta \lambda +\gamma   \end{equation}
In this work, we employed the parameter set as ($\alpha , \beta , \gamma $) = (4.82, 17.33, 221.59) for slit spectroscopy.
For an object not on the slit (for example, a standard star in the exposed region described in Section \ref{cal}), 
position can be corrected by adding a position offset between the slit position and the target object taking account of the distortion.

\subsection{Absolute calibration}\label{cal}
To obtain a response function used for converting from ADU unit to physical unit in surface brightness, we used a spectrum of a standard star KF09T1, a K0III star with m$_H$=8.114 mag.
This star is also used for the absolute calibration of the Infrared Array Camera (IRAC) on Spitzer \citep{Reach2005}.

\begin{figure*}
  \begin{center}
    \FigureFile(160mm,100mm){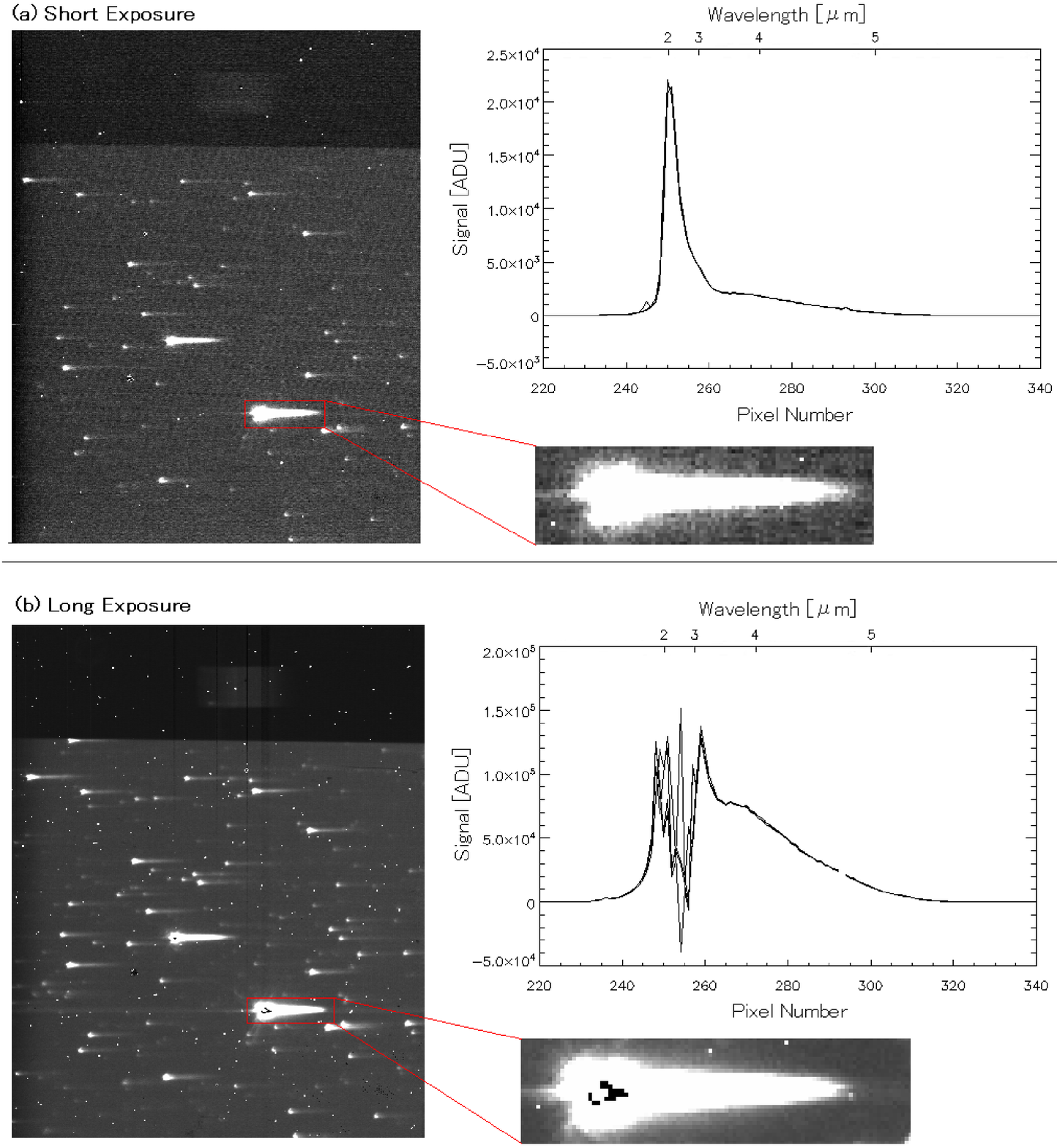}
  \end{center}
  \caption{Images of the standard star KF09T1 with (a) short exposure and (b) long exposure. Zoom-in images of KF09T1 and its spectrum are also shown.
               Since short wavelength region ($x$=245-260) of the KF09T1 spectrum with long exposure was saturated, spectrum with short exposure was combined to derive the response function.}
  \label{calstar}
\end{figure*}

Figure \ref{calstar} shows the images of KF09T1. 
AKARI data include both long exposure images and short exposure images, and we used the long exposure images for diffuse sky spectra in this study.
However, since short wavelength region of the KF09T1 spectrum is saturated or non-linearly accumulated with long exposure as shown in Figure \ref{calstar},
we also used the short exposure images to derive the spectrum of KF09T1 in ADU unit.
The long/short exposure ratio is known to be 38.0 in ADU\footnote{The ratio in the integration time is 9.5 and the difference in a data compression factor is 4.0.} and this factor is used for converting the response function from short exposure to long exposure.
Since the spectrum of KF09T1 is provided with 3\% accuracy and the solid angle equivalent to the slit width of IRC is also known,
the response function from ADU to surface brightness unit is obtained by comparing these spectra.

\begin{figure}
  \begin{center}
    \FigureFile(80mm,50mm){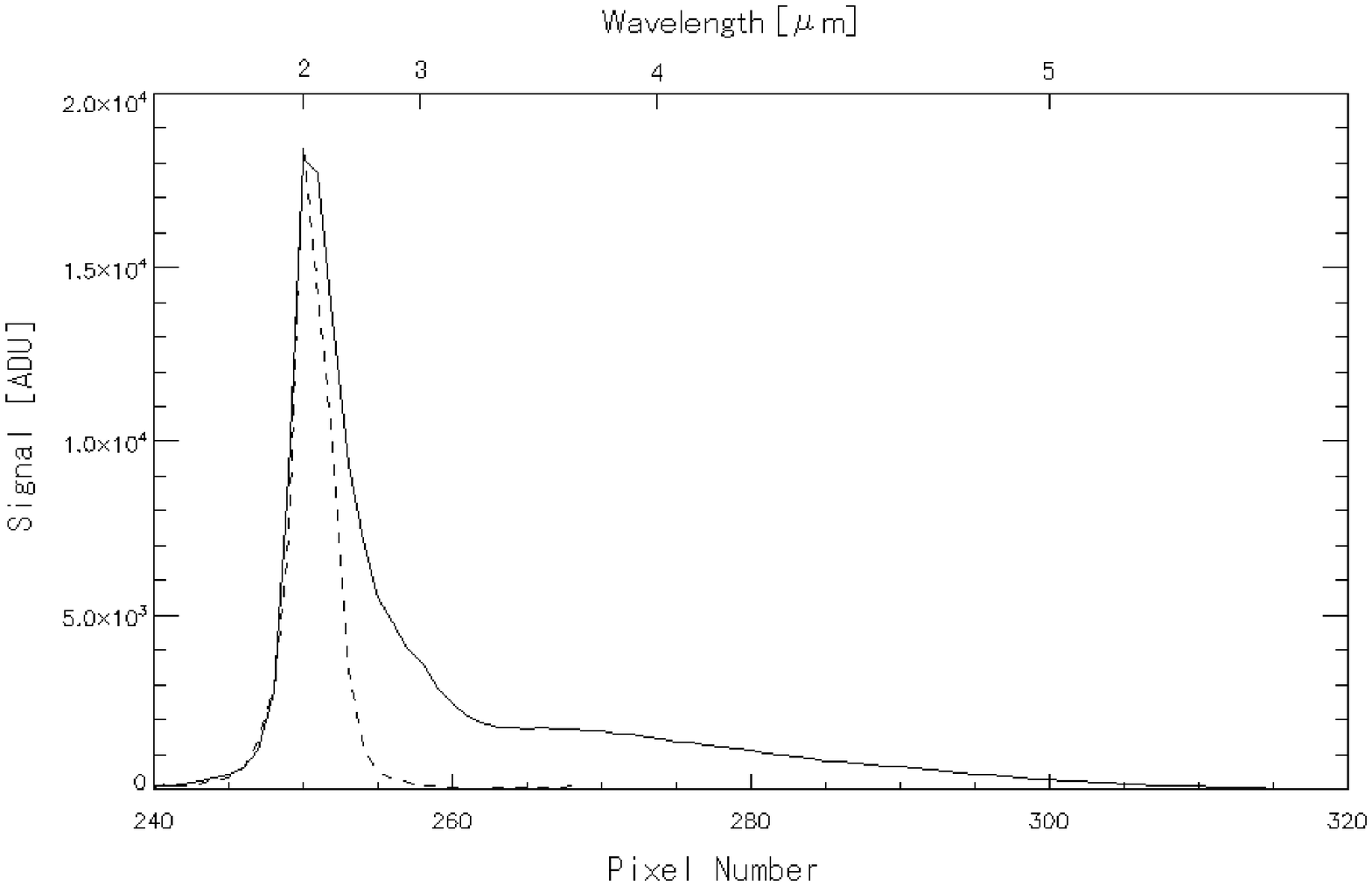}
  \end{center}
  \caption{Spectrum of KF09T1 with short exposure (solid line) and its best fit photometry at shortest wavelengths with IRC N2 PSF profile (broken line).} 
  \label{PSFfit}
\end{figure}

\begin{figure}
  \begin{center}
    \FigureFile(80mm,50mm){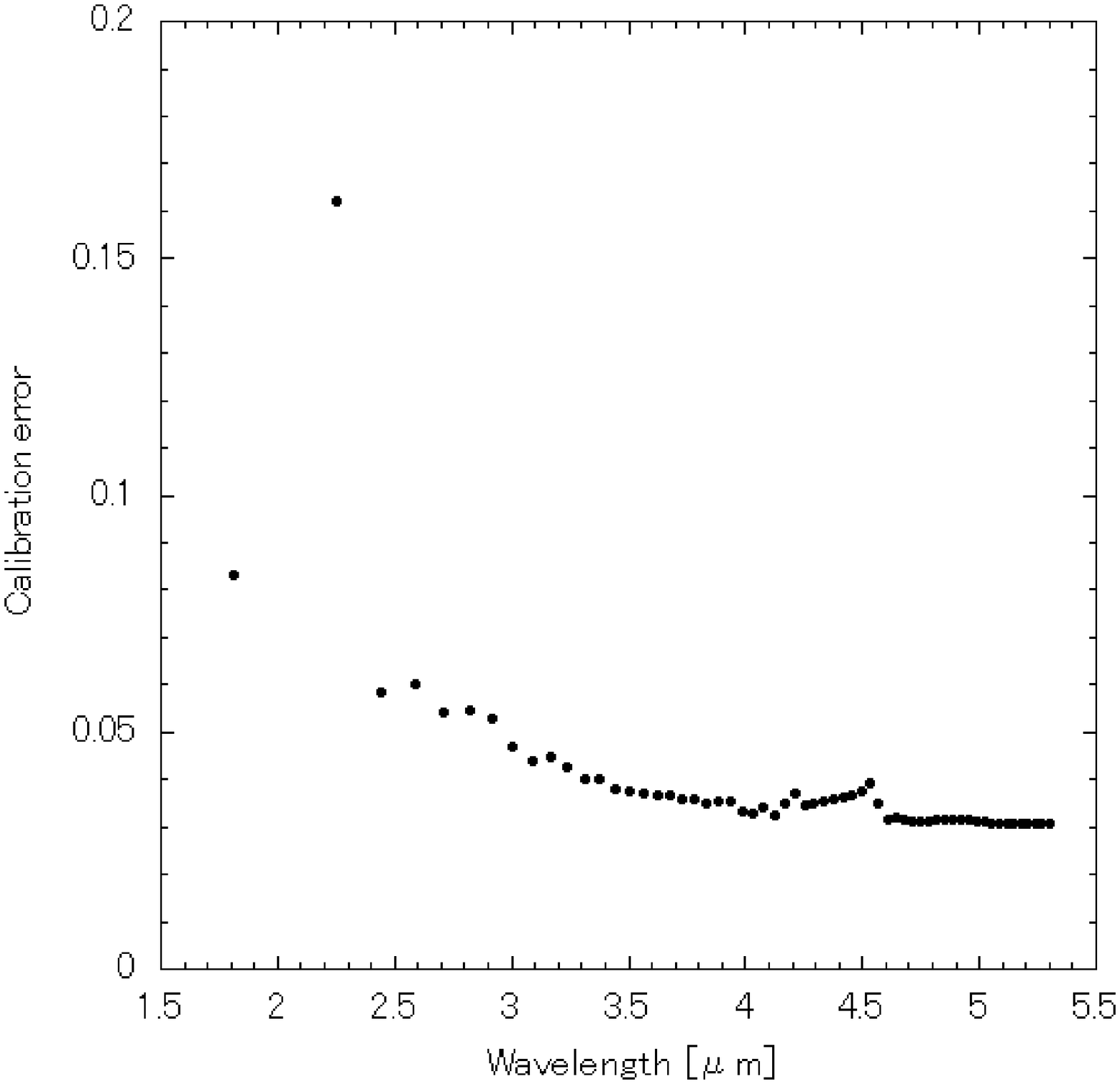}
  \end{center}
  \caption{The total systematic calibration error owing to 3\% accuracy of the spectrum of the standard star and the sub-pixel misalignment.} 
  \label{calerror}
\end{figure}

The other factor of calibration error comes from wavelength uncertainty at given pixels owing to the sub-pixel misalignment of the star,
since it is difficult to estimate the exact position of the star on the detector array within sub-pixel accuracy.
Therefore, this type of calibration error was estimated by wavelength shift equivalent to the sub-pixel position uncertainty of $\pm $0.5 pixels.
As shown in Figures \ref{mask} and \ref{baseline}, photons converge at the shortest wavelengths owing to wavelength dispersion of the prism, 
i.e. wavelength width $\Delta \lambda $ at the pixels of shortest wavelengths is much greater than other pixels.
Therefore, small sub-pixel position uncertainty causes a large systematic calibration error at the shortest wavelengths ($<$2.2 $\mu$m).
In addition, since the signal at the shortest wavelength region is strong, its PSF wing contaminates neighbor pixels, which is another error source. 
To avoid them, a photometric method was employed only for the shortest wavelength by using the PSF profile of IRC imaging data with N2 (1.9-2.8 $\mu$m) filter.
Figure \ref{PSFfit} is the spectrum of KF09T1 with short exposure and its best fit photometry at shortest wavelengths with the N2 PSF profile,
and it shows that shortest-side wing of the spectrum can be fit by the N2 PSF well.
For deriving the absolute response of the diffuse sky, N2 PSF fitting at shortest wavelengths was conducted and surface brightness of the sky at shortest wavelength was obtained by comparing the best-fit PSF of KF09T1 at first.
Then, spectrum at other wavelengths were derived after subtracting the best-fit PSF profiles.
The total systematic calibration error is shown in Figure \ref{calerror}, which is larger than the statistic error at $<$3 $\mu$m.

\subsection{Comparison to the previous data}
\begin{figure}
  \begin{center}
    \FigureFile(80mm,50mm){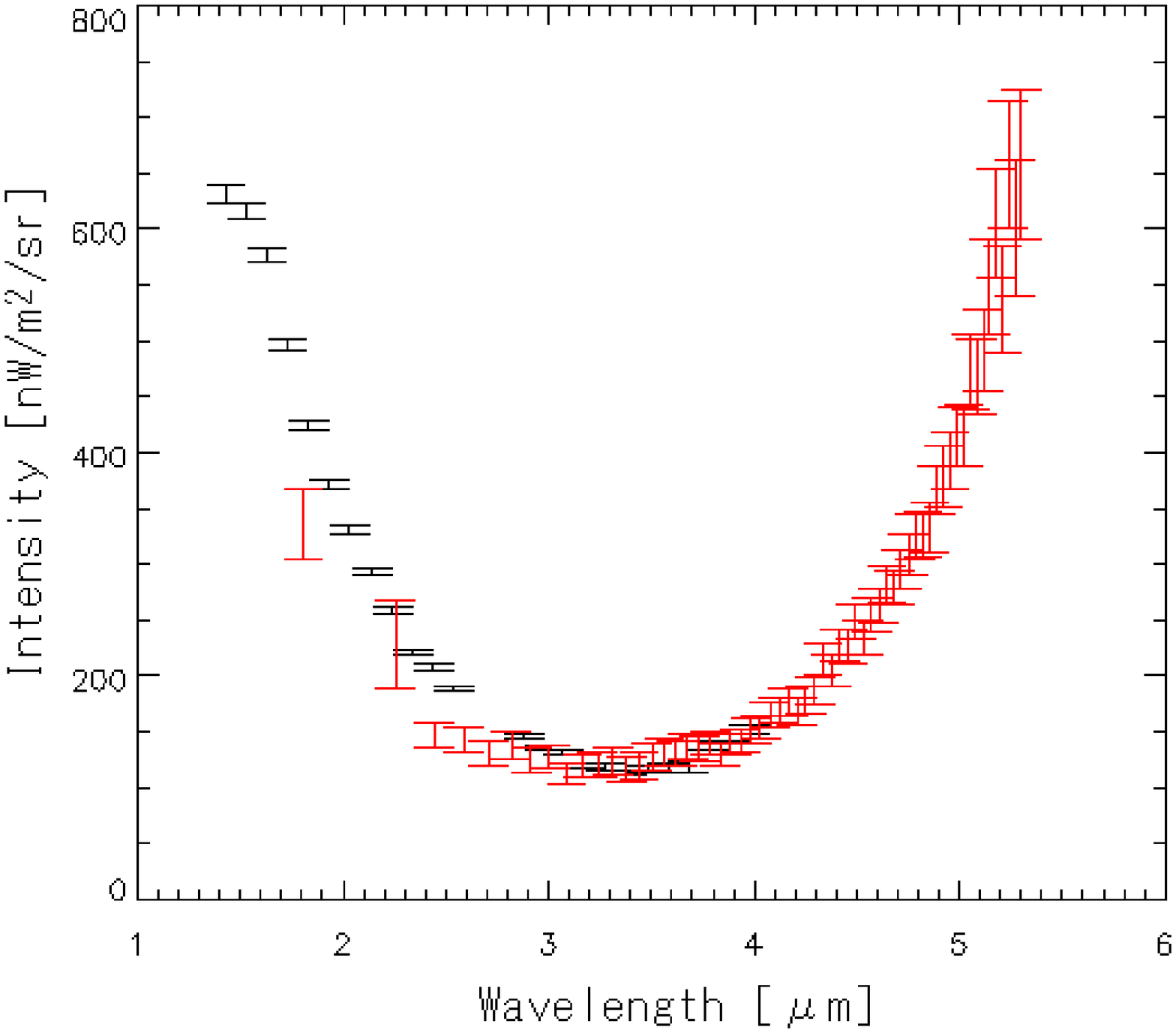}
  \end{center}
  \caption{Diffuse sky spectrum at ecliptic coordinate ($\lambda $, $\beta $) = (130$^{\circ}$, 20$^{\circ}$) with IRTS (black, \citet{Matsumoto2013}) and AKARI IRC (red, this work) after the seasonal variation correction.
  Contribution of unresolved faint stars in IRTS spectrum is subtracted by a model.} 
  \label{IRTS-AKARI}
\end{figure}

Data reduction described here was applied to all the data, and the spectral catalog of diffuse sky was obtained.
To check the data quality of these spectra, we compared them with the diffuse sky spectra at NIR observed with IRTS \citep{Matsumoto2013}.
Contribution of unresolved faint stars in the IRTS spectra is subtracted by the SKY model for the Galactic point sources \citep{Cohen1994}. 
Direct comparison was done by using the spectra at the same fields observed by both IRTS and AKARI.
Figure \ref{IRTS-AKARI} is an example of the direct comparison of the diffuse sky spectra at ecliptic coordinate ($\lambda $, $\beta $) = (130$^{\circ}$, 20$^{\circ}$)
taken with IRTS in April 1995 and with AKARI IRC in December 2006 after the seasonal variation correction of ZL using the DIRBE ZL model \citep{Kelsall98},
showing that the spectral shape of the sky is consistent with each other.

In addition, the absolute sky brightness at NEP in our dataset is also consistent with that from AKARI IRC imaging data of 114 \nw\ at 2.4 $\mu$m, 73 \nw\ at 3.2 $\mu$m, and 105 \nw\ at 4.1 $\mu$m \citep{Matsumoto2011},
those are obtained from the dataset in the same period with higher sensitivity for point source removal at foreground.

\section{Zodiacal Light Spectrum}
\subsection{Spectral shape}
Since the AKARI data in this study have wide range of ecliptic latitudes and seasons, we can test the spectral shape dependence at various ecliptic latitudes and seasons.
DGL was subtracted by the template DGL spectrum and the correlation between 3.3 $\mu$m PAH band emission and 100 $\mu$m thermal emission \citep{Schlegel1998} described in Paper II.
Since the 3.3 $\mu$m PAH band was not detected at high Galactic and low ecliptic latitude regions, we conclude that the PAH band emission does not contribute to ZL.

\begin{figure*}
  \begin{center}
    \FigureFile(160mm,100mm){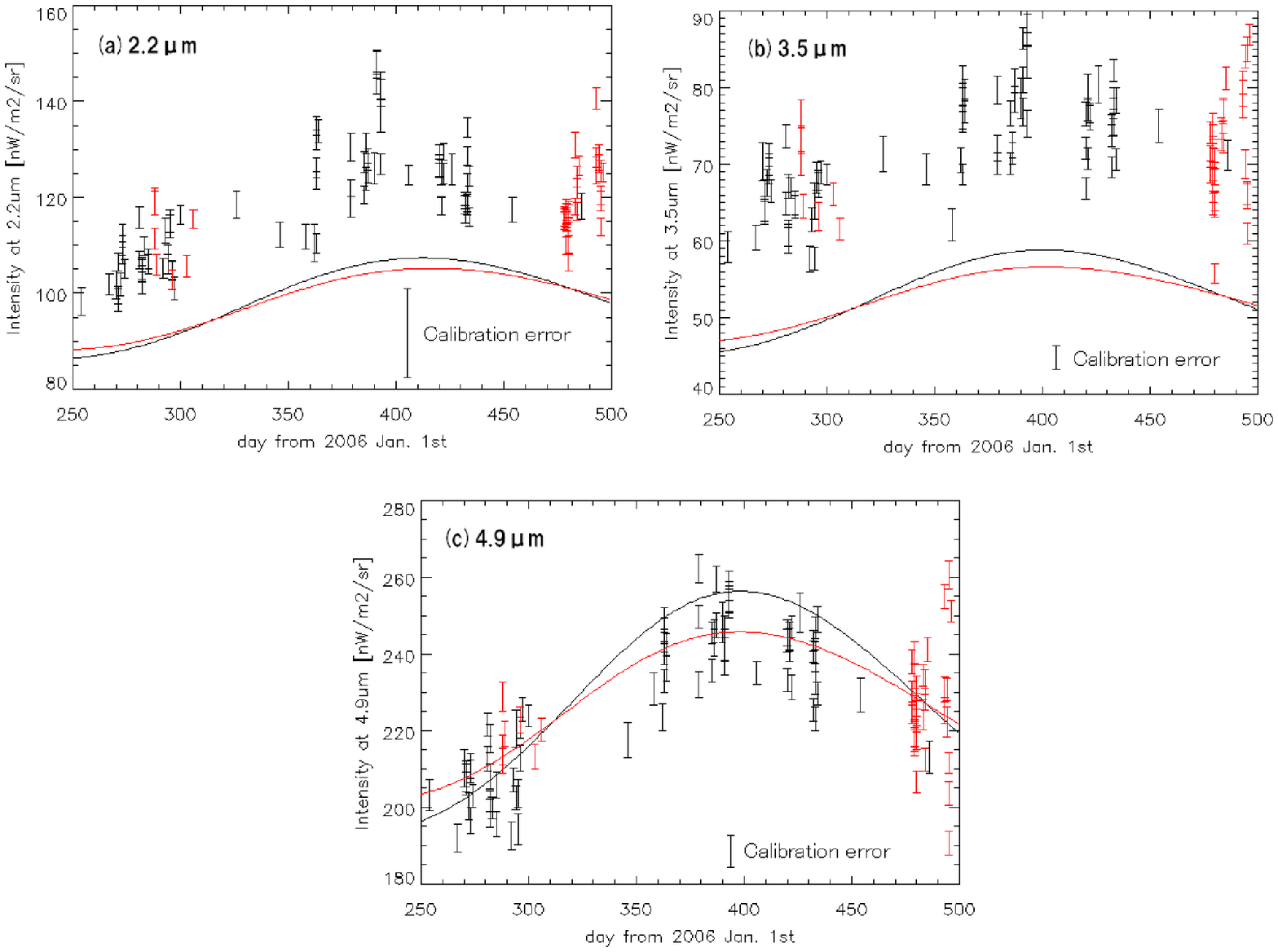}
  \end{center}
  \caption{Seasonal variation of the diffuse sky brightness taken in the NEP fields at (a) 2.2 $\mu$m, (b) 3.5 $\mu$m, and (b) 4.9 $\mu$m. 
  Black shows the data at AKARI NEP field, and red shows at IRAC dark field.
  Error bars shows only the statistic errors and the calibration errors are shown separately in each figure. 
           The solid lines show the model brightness from the DIRBE ZL model \citep{Kelsall98}.}
  \label{seasonal}
\end{figure*}

\begin{figure*}
  \begin{center}
    \FigureFile(160mm,100mm){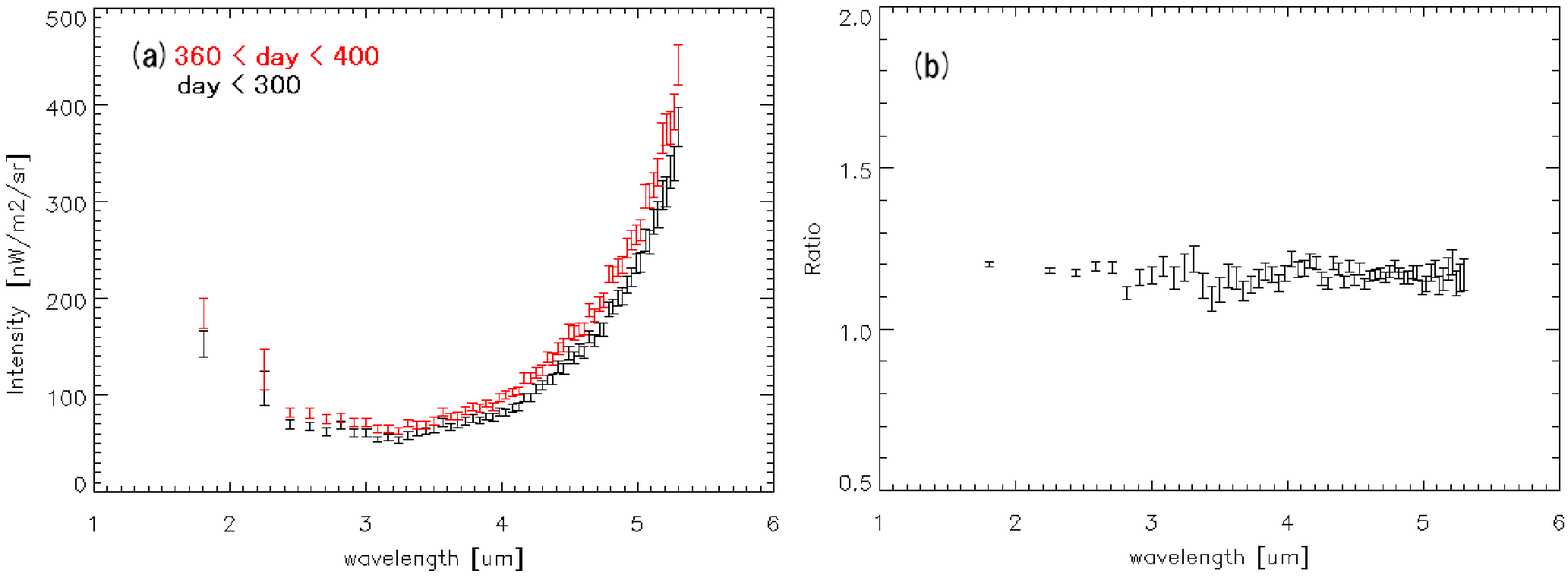}
  \end{center}
  \caption{Seasonal variation of the diffuse sky spectrum dominated by ZL at NEP.
               In Figure (a), sky spectra averaged between day 250 and 300 (black) and between day 360 and 400 (red) are shown, where the day is counted from January 1st 2006.
               Figure (b) shows the ratio between the spectra in Figure (a).
               Error bars in Figure (a) consist of both the calibration error and the statistic error, while error bars in Figure (b) consist only the statistic error.} 
  \label{difference_season}
\end{figure*}

Because of the AKARI's Sun-synchronous polar orbit, the NEP field was observed frequently and allowed us to monitor the seasonal variation of ZL.
Seasonal variation of ZL detected by the IRC imaging data is reported \citep{Pyo2010, Pyo2012}, and here we report the seasonal variation from the spectral data.
Figure \ref{seasonal} compares the seasonal variation observed by AKARI IRC with the DIRBE ZL model at three wavelength bands.
The excesses of observed brightness from the model brightness at 2.2 $\mu$m and 3.5 $\mu$m are mainly caused by the isotropic EBL reported by \citet{Hauser98}, \citet{Matsumoto05}, \citet{Matsumoto2013}, and Paper III.
Figure \ref{difference_season} shows the spectra of the maximum and minimum cases due to the seasonal variation at NEP and its ratio.
As shown in this figure, the spectral shape of ZL is constant within $<2.5$ \% at NEP and only the brightness of ZL changes depending on the season.

\begin{figure*}
  \begin{center}
    \FigureFile(160mm,100mm){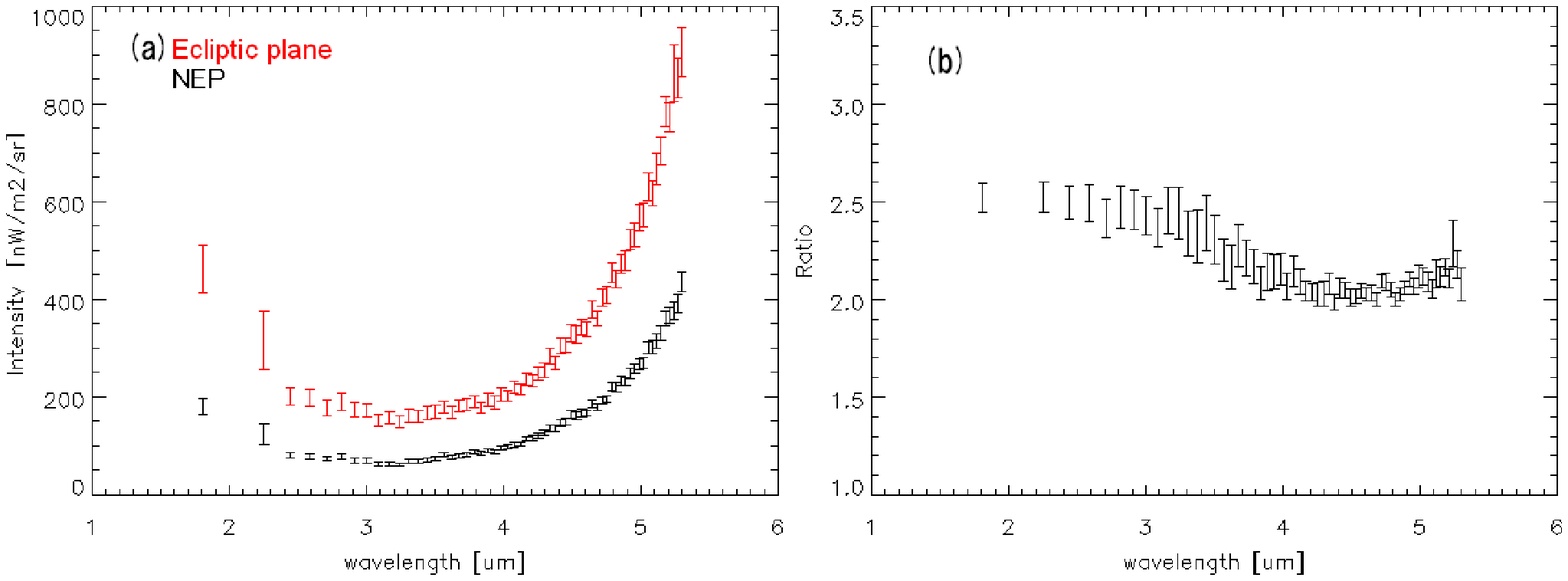}
  \end{center}
  \caption{Ecliptic latitude dependence of the diffuse sky spectrum dominated by ZL.
               Averaged sky spectrum at NEP (black) and ecliptic plane (red) are shown in Figure (a).
               Figure (b) shows the ratio between the spectra in Figure (a).
               Error bars in Figure (a) consist of both the calibration error and the statistic error, while error bars in Figure (b) consist only the statistic error.} 
  \label{difference_EclLat}
\end{figure*}

Averaged NEP spectrum was derived from the spectra at the ZL brightest season (day 360-400), because seasonal difference of ZL within this period can be negligible.
Figure \ref{difference_EclLat} compares such the averaged NEP spectrum with the averaged spectrum at the ecliptic plane.
The ratio of these spectra has a step at around 3.5 $\mu$m where the scattered and thermal emission component are comparable,
and both sides of the step are flat.
It can be understood with a simple explanation; spectral shapes of the scattered components and thermal emission component do not depend on location,
which make the flat ratios, but the relative brightness between those components vary with location, which makes the step at around 3.5 $\mu$m.

\subsection{High color temperature and sub-micron particles}\label{submicron}
The shape of the IPD smooth cloud is flattened and temperature of IPD decrease with increasing distance from the Sun.
For example, $T(r)=286[K]\cdot r^{-0.467}$ is employed in the DIRBE ZL model \citep{Kelsall98} where $r$ is the distance from the Sun in au.
Therefore, the color temperature of ZE at ecliptic pole becomes higher than that at the ecliptic plane,
because closer dusts with higher temperature contributes to ZE more than that at the ecliptic plane. 
For example, IRAS observation at 12, 25, and 60 $\mu$m obtained $275 \pm 57$ K at the ecliptic pole and $244 \pm 44$ K at the ecliptic plane\footnote{100 $\mu$m data were also used at the ecliptic plane.} with single temperature fitting \citep{Hauser1984},
and ISO spectroscopic observation at 5-16 $\mu$m obtained $274.0 \pm 1.1$ K at the ecliptic pole,
while $268.5 \pm 0.4$ K (solar elongation = 60$^{\circ}$) and $244.1 \pm 0.6$ K (solar elongation = 120$^{\circ}$) at the ecliptic plane \citep{Reach03}.
A similar result can be obtained also from the DIRBE ZL model \citep{Kelsall98}.
This temperature is named as "peak temperature" in this paper, since it was derived at around the peak wavelength ($>$5 $\mu$m) of ZE spectrum.

\begin{figure}
  \begin{center}
    \FigureFile(80mm,50mm){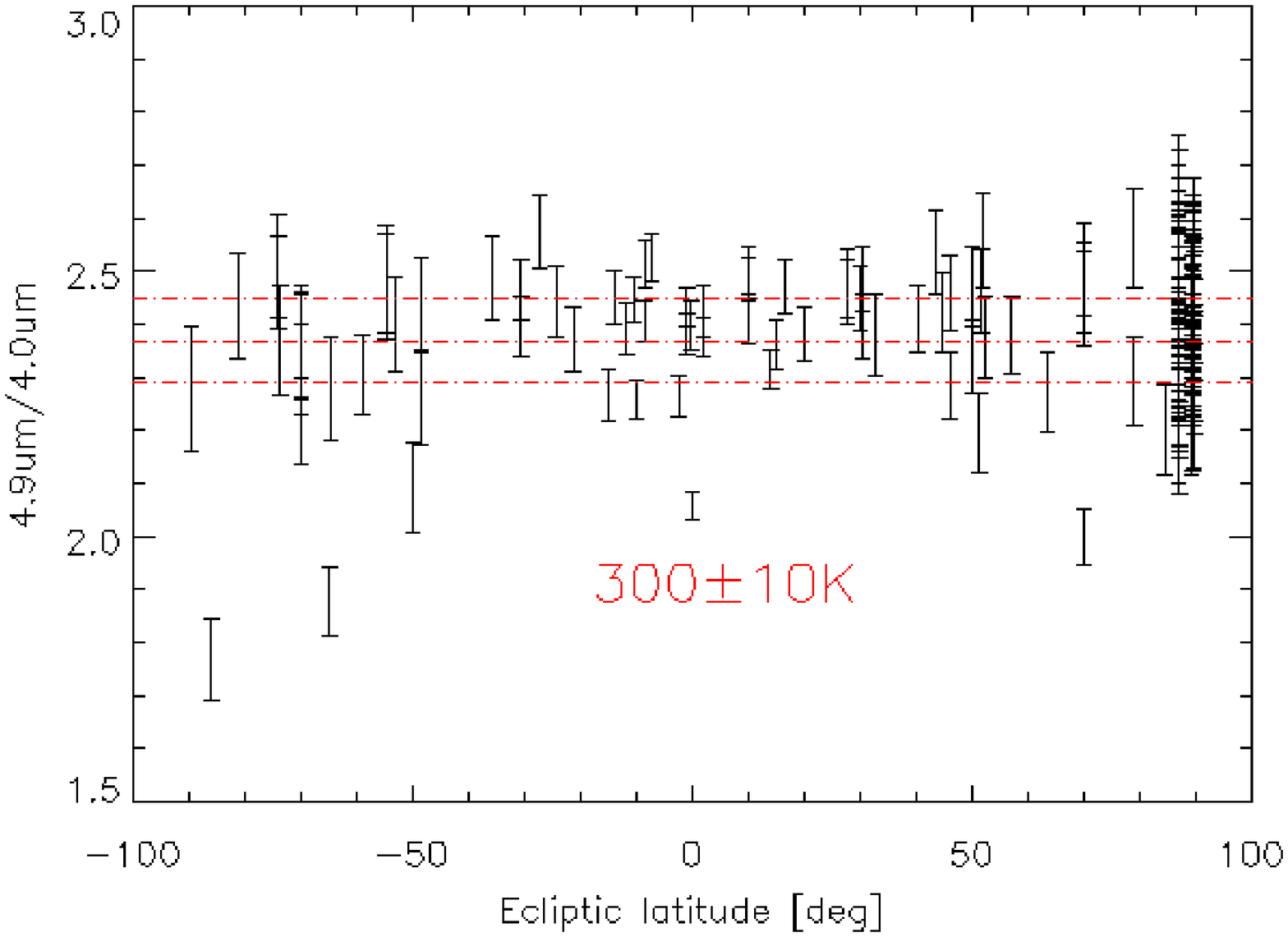}
  \end{center}
  \caption{Color between 4.0 $\mu$m and 4.9 $\mu$m as a function of ecliptic latitude.
           This plot shows the color temperature of ZE obtained with AKARI does not show any dependence on ecliptic latitude contrary to observations with IRAS, ISO and DIRBE,
           and the obtained color temperature of 300$\pm $10 K is higher than the estimated temperature at the longer wavelengths.}
  \label{temperature}
\end{figure}

We tested this temperature dependence at various ecliptic latitude in our AKARI dataset of diffuse spectra at 3-5 $\mu$m wavelength region,
while the temperature in previous studies (IRAS, ISO, and DIRBE) were determined at longer wavelengths ($>$5 $\mu$m) of ZE spectrum.
Contrary to previous observations, no difference in color temperature of the ZE spectrum was detected with AKARI IRC in this study.
Figure \ref{temperature} shows the brightness ratio between 4.0 $\mu$m and 4.9 $\mu$m as a function of ecliptic latitude.
These wavelengths are chosen for avoiding the contamination from scattering component at shorter wavelengths and the 3.3 $\mu$m PAH band.
Although the ratio between 4.0 $\mu$m and 4.9 $\mu$m directly correspond to the color temperature of ZE, 
any dependence on ecliptic latitude cannot be found and the color temperature of ZE is 300$\pm $10 K at any ecliptic latitude.
Similar analysis was conducted at other wavelengths but no ecliptic latitude dependence was found from this AKARI ZL data.
In conclusion, the spectral shape of ZE at 3-5 $\mu$m with AKARI IRC does not show any dependance on ecliptic latitude, and its color temperature (300 K) is higher than the peak temperature.

In fact, such higher color temperature was also detected by the previous IRTS measurement of ZL \citep{Ootsubo98, Ootsubo2000}, 
in which 266 K was employed for 6-12 $\mu$m and 300K was employed for 3-6 $\mu$m, 
meaning that both AKARI and IRTS results show that color temperature at shorter wavelength (3-5 $\mu$m) is higher than the peak temperature ($\sim$260 K) derived at $>$5 $\mu$m.
In addition, IRTS result also showed that the spectral shape does not change apparently with the ecliptic latitude \citep{Ootsubo98}, 
which is also consistent with our AKARI result.
\citet{Hong2009} also analyzed the composite spectral energy distribution of the ZE from the IRTS, COBE/DIRBE, and AKARI observations.  
They suggested two components for the IPD cloud model, each of which has the dust temperature of $\sim$243 K and $\sim$386 K at 1 au from the Sun.

Similar temperature difference between 3-4 $\mu$m and $>$10 $\mu$m was observed in the dust grains associated with outburst of Jupiter-family comet 17P/Holmes in 2007.
A NIR spectroscopy showed that color temperature around 3-4 $\mu$m was 360 $\pm$ 40 K \citep{Yang2009}, 
while a MIR observation showed that color temperature between 12.4 $\mu$m and 24.5 $\mu$m was $\sim$200 K \citep{Watanabe2009}.
\citet{Ishiguro2010} explained this temperature difference by the heterogeneity in particle size, that is, the hotter component consists of sub-micron absorbing particles whereas the colder component consists of large particles ($>1\ \mu$m).  
In the same context, the higher color temperature of ZE detected in our study can be an evidence of existence of sub-micron dust particles.
\citet{Reach1988} also shows that sub-micron particles, or graphite particles with $>100$ $\mu$m size, are required to be $>300$ K at 1 au from the Sun.  
The existence of sub-micron size particles in IPD was confirmed as microcraters on the dust particles recovered from the near-Earth asteroid 25143 Itokawa by the HAYABUSA spacecraft \citep{Nakamura2012}.
On the other hand, the color of ZL in the scattered sunlight regime was observed to be redder than the solar spectrum \citep{Matsuura95, Matsumoto96, Tsumura10}, 
implying that dust particles larger than 1 $\mu$m are mainly responsible for the scattered component in ZL \citep{Matsuura95}.
This large particles produces the ZE with the peak temperature of $\sim$260 K at $>10\ \mu$m.
Since the ZE spectrum is dominated by the nearby dust from the Earth with higher temperature, a lack of color temperature variation toward ecliptic latitude can be understood
if the sub-micron particles contributing the higher color temperature locate mainly within the scale height of the IPD cloud ($\sim$0.2 au).

\subsection{Modeling of the zodiacal light}
\begin{figure}
  \begin{center}
    \FigureFile(80mm,50mm){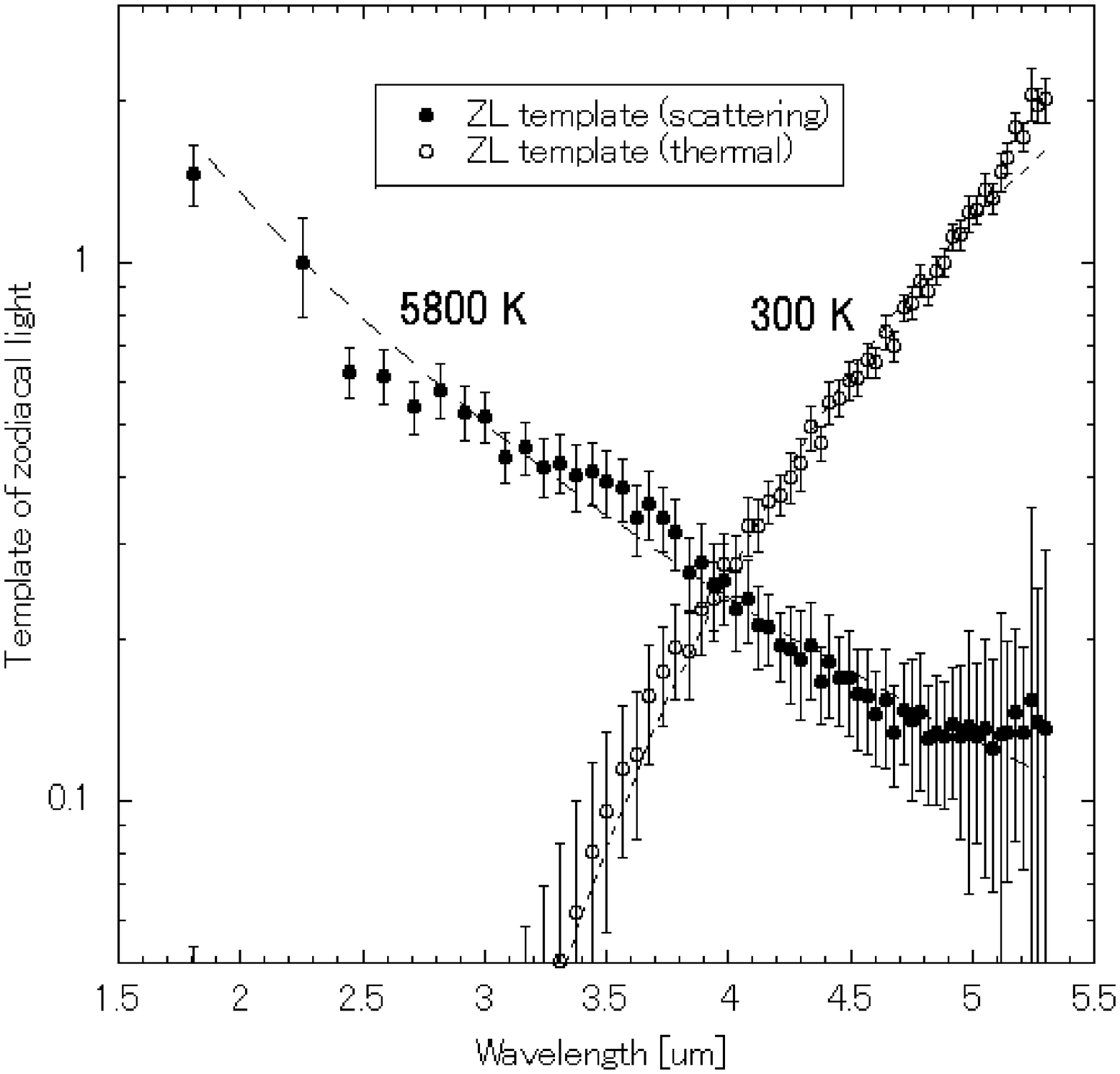}
  \end{center}
  \caption{
  The scattered sunlight component (solid circle) normalized at 2.2 $\mu$m ($ZL_{\textrm{temp}}^{\textrm{scat}}(\lambda )$) 
  and the thermal emission component (open circle) normalized at 4.9 $\mu$m ($ZL_{\textrm{temp}}^{\textrm{thermal}}(\lambda )$). 
  These spectra are prorated from the differential spectrum by two temperature fitting of 5800 K and 300 K.} 
  \label{ZL_temp}
\end{figure}

The spectral shape of ZL can be separated from the the isotropic components such as EBL by differencing the spectra between NEP and the ecliptic plane \citep{Matsumoto96, Tsumura10}.
The differential spectrum can be explained well by the combination of black body emissions of 5800 K (scattered solar light) and 300 K (IPD thermal emission).
Since the relative brightness between the scattering component and the thermal emission component vary at position as shown in  Figure \ref{difference_EclLat} (d),
these two components should be evaluated separately.
The separated spectra of the scattered sunlight and thermal emission were prorated from the differential spectrum by the two temperature (5800 K and 300 K) fitting.
These separated spectra of the scattered component normalized at 2.2 $\mu$m ($ZL_{\textrm{temp}}^{\textrm{scat}}(\lambda )$)
and the thermal emission component normalized at 4.9 $\mu$m ($ZL_{\textrm{temp}}^{\textrm{thermal}}(\lambda )$) are shown in Figure \ref{ZL_temp}.
By using these two spectra as template spectra of ZL, we can estimate a local ZL spectrum at any position $i$ by scaling these templates to the DIRBE ZL model at
2.2 $\mu$m and 4.9 $\mu$m, i.e.;
\begin{equation}  ZL_i(\lambda ) =  ZL_i^{\textrm{scat}}(\lambda ) + ZL_i^{\textrm{thermal}}(\lambda ) \label{eq_ZL} \end{equation}
\begin{equation}  ZL_i^{\textrm{scat}}(\lambda ) =  DIRBE_i^{2.2 \mu m} \cdot ZL_{\textrm{temp}}^{\textrm{scat}}(\lambda ) \end{equation}
\begin{eqnarray}
ZL_i^{\textrm{thermal}}(\lambda )  =    \nonumber  \\ 
& \hspace{-23mm}  [DIRBE_i^{4.9 \mu m} - ZL_i^{\textrm{scat}}(4.9 \mu m)]  \cdot ZL_{\textrm{temp}}^{\textrm{thermal}}(\lambda )  
\end{eqnarray}

\begin{figure*}
  \begin{center}
    \FigureFile(160mm,100mm){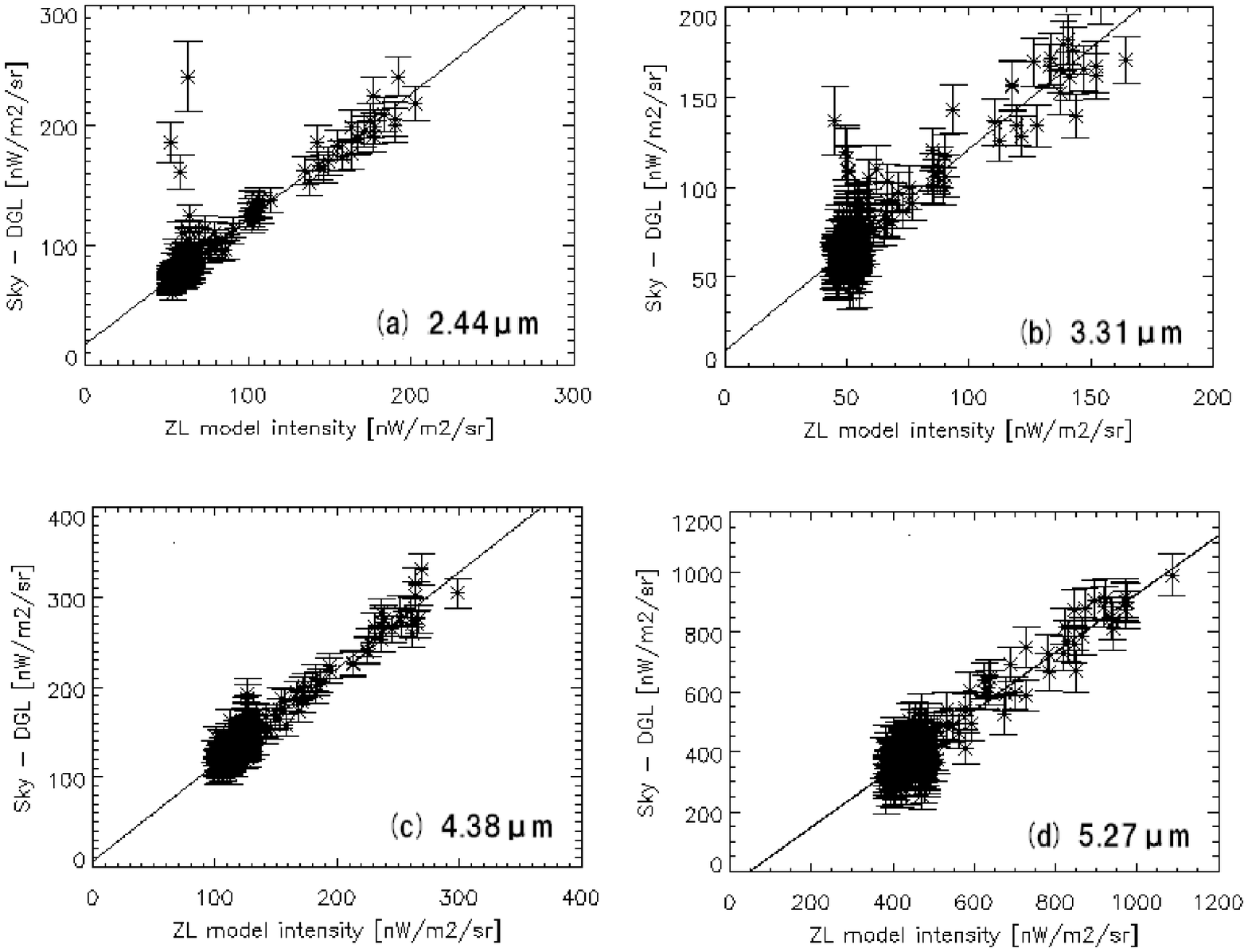}
  \end{center}
  \caption{Examples of correlation between $SKY_i(\lambda ) - DGL_i(\lambda )$ and $ZL_i(\lambda )$ from equation (\ref{eq_ZL}) at (a) 2.44 $\mu$m, (b) 3.31 $\mu$m, (c) 4.38 $\mu$m, and (d) 5,27 $\mu$m.
               The best fit lines after 3$\sigma $ clipping for outlier rejection are also shown}
  \label{correlation}
\end{figure*}

\begin{figure}
  \begin{center}
    \FigureFile(80mm,50mm){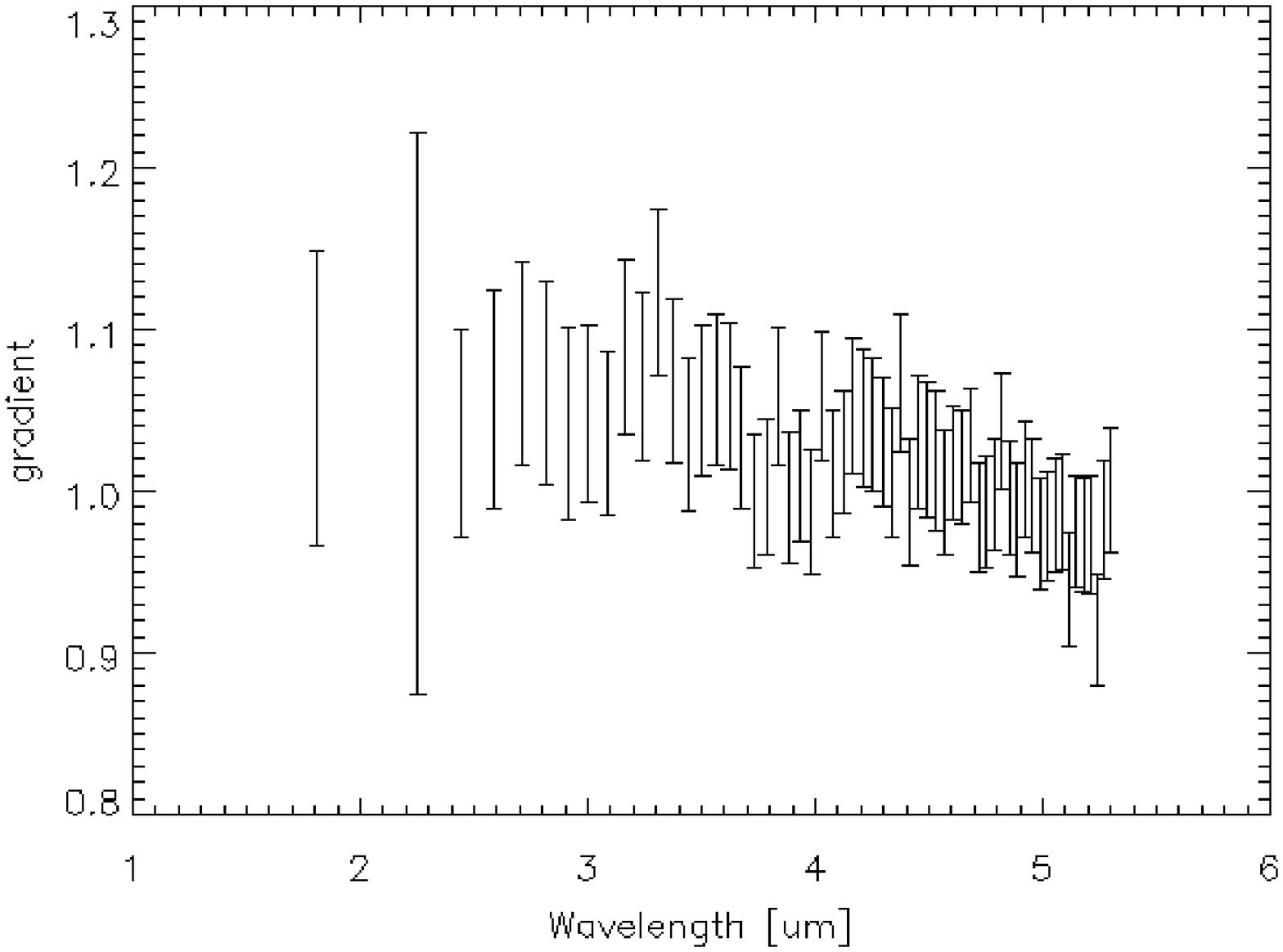}
  \end{center}
  \caption{Gradients of the correlation between $SKY_i(\lambda ) - DGL_i(\lambda )$ and $ZL_i(\lambda )$ shown in Figure \ref{correlation}.
  This is used as a correction factor of ZL, $C(\lambda)$.
  The large error bars are dominated by the calibration error shown in Figure \ref{calerror}.}
  \label{ZL_correction}
\end{figure}

The derived ZL spectrum is tested by a correlation analysis.
The diffuse sky spectrum includes ZL, DGL, and EBL, i.e.
\begin{equation} SKY_i (\lambda ) = ZL_i (\lambda ) + DGL_i (\lambda ) +EBL_i(\lambda )  \end{equation}
Thus, we checked the correlation between the DGL subtracted spectra $SKY_i(\lambda ) - DGL_i(\lambda )$ and the derived ZL spectra from equation (\ref{eq_ZL}).
In this analysis, spectra at low Galactic latitude ($\mid b \mid <30^{\circ}$) were not used to avoid the contamination from the PAH emission at 3.3 $\mu$m.
For each wavelength, the correlation between $SKY_i(\lambda ) - DGL_i(\lambda )$ and $ZL_i(\lambda )$ from the equation (\ref{eq_ZL}) was obtained as shown in Figure \ref{correlation}.
Best fit lines in Figure \ref{correlation} were obtained after 3$\sigma $ clipping for outlier rejection, and their gradients  $C(\lambda)$ are shown in Figure \ref{ZL_correction}.
Fairly good correlations can be seen and the gradient $C(\lambda)$ is basically consistent with unity within the error bars, indicating that the DIRBE ZL model is consistent with our data set.
However, the gradients at short wavelengths ($<$3.5 $\mu$m) are larger than unity by $\sim$5\%, indicating that the DIRBE ZL model underestimates the ZL surface brightness in these wavelengths.
This trend is consistent with the fact that the estimated ZL brightness by the DIRBE ZL model are smaller than other models by 5-8 \% at 1.25 $\mu$m and 11-14 \% at 2.2 $\mu$m \citep{Kelsall98}.

\begin{figure}
  \begin{center}
    \FigureFile(80mm,50mm){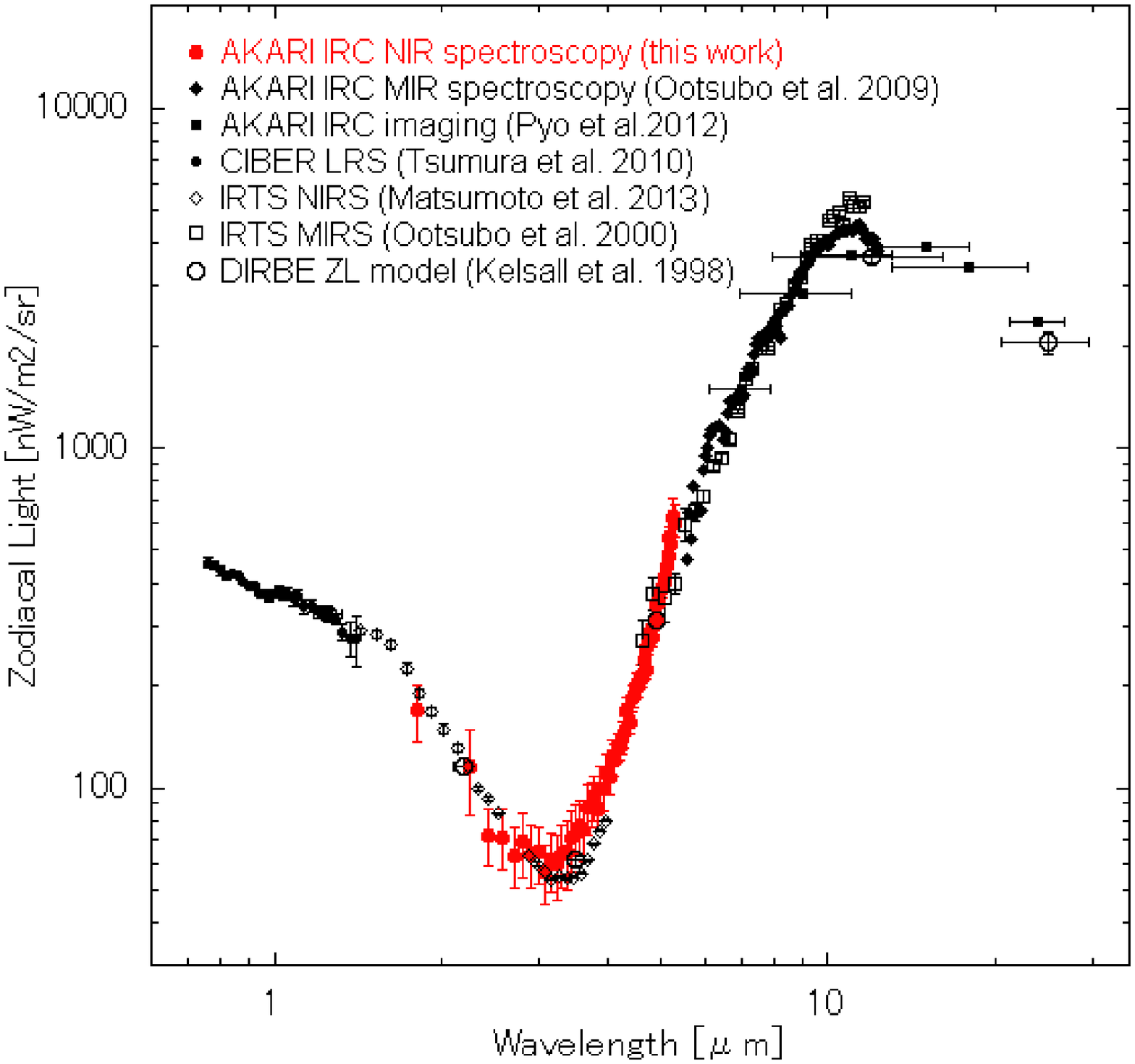}
  \end{center}
  \caption{ZL spectrum at a high Galactic and ecliptic latitude region obtained by various space-based instruments. 
                 Red points show our AKARI/IRC data derived from the equation (\ref{eq_ZL2}) in this paper. 
                 All these data are normalized to the data from DIRBE ZL model.}
  \label{ZL}
\end{figure}

The ZL spectrum in equation (\ref{eq_ZL}) is corrected by this obtained gradients $C(\lambda)$ like this
\begin{equation}  ZL_i^{corr}(\lambda ) =  C(\lambda) \cdot  ZL_i(\lambda )  \label{eq_ZL2} \end{equation}
Figure \ref{ZL} shows the ZL spectrum at a high Galactic and ecliptic latitude region derived from the equation (\ref{eq_ZL2}) with the ZL intensities by privies studies normalized by the DIRBE ZL model,
showing a good agreement of our ZL estimation to the others.

\section{Conclusion}
We analyzed low-resolution spectra of ZL at 1.8-5.3 $\mu$m with wavelength resolution of $\lambda /\Delta \lambda \sim 20$ from the AKARI IRC diffuse spectral catalog.
ZL spectrum has two components, scattered sunlight component ($<$3 $\mu$m) and thermal emission component ($>$3 $\mu$m),
and the spectral shape of each component does not show any dependance of location and season while their brightness ratio depends on the ecliptic coordinates.
The color temperature obtained from our ZL spectra at 3-5 $\mu$m is 300$\pm$10 K at any ecliptic lattitude.
This temperature is higher than the temperature obtained by longer wavelength region ($>$5 $\mu$m),
and this can be an evidence of sub-micron particles in IPD around the Earth.
We also construct a model to estimate the ZL spectrum by using template ZL spectra obtained from our spectral dataset and the DIRBE ZL model.

\bigskip
This research is based on observations with AKARI, a JAXA project with the participation of ESA.
This research is also based on significant contributions of the IRC team.
We thank Dr. Ootsubo Takafumi (Tohoku university) for providing us the ZL data by IRTS/MIRS and IRC MIR spectroscopy, 
Prof. Ishiguro Masateru (Seoul National University) and Dr. Sarugaku Yuki (ISAS/JAXA) for discussion about IPD,
Dr. Usui Fumihiko (ISAS/JAXA) and Dr. Yamamura Issei (ISAS/JAXA) for preparation of data release,
and Dr. Egusa Fumi (ISAS/JAXA) and Mr. Arimatsu Ko (ISAS/JAXA) for providing information about the data reduction.
The authors acknowledge support from Japan Society for the Promotion of Science, KAKENHI (grant number 21111004 and 24111717).

\end{document}